\documentclass[a4paper, 11pt]{article}

\usepackage[english]{babel}
\usepackage[utf8x]{inputenc}
\usepackage[T1]{fontenc}

\usepackage[a4paper,top=2cm,bottom=2cm,left=2cm,right=2cm,marginparwidth=1.75cm]{geometry}
\usepackage{amsmath}
\usepackage{amssymb}
\usepackage{multirow}
\usepackage{graphicx}
\usepackage{natbib}
\usepackage{wasysym} 
\usepackage{subcaption}
\usepackage[colorinlistoftodos]{todonotes}
\PassOptionsToPackage{hyphens}{url}\usepackage[colorlinks=true, allcolors=red]{hyperref}
\hypersetup{breaklinks=true}

\usepackage{lscape}

\usepackage{enotez}

\let\footnote=\endnote

\newcommand{\papertitle}{{The gig economy in Poland: evidence based on mobile big data}} 

\title{\papertitle}
\author{Beręsewicz Maciej
\footnote{Corresponding author: \href{mailto:maciej.beresewicz@ue.poznan.pl}{maciej.beresewicz@ue.poznan.pl}. Poznań University of Economics and Business, Poland; Statistical Office in Poznań, Poland. \textbf{Acknowledgements}: The study was conducted as part of the research project \textit{Economics in the face of the New Economy}, financed under the Regional Initiative for Excellence programme of the Minister of Science and Higher Education of Poland, years 2019-2022, grant no. 004/RID/2018/19, financing 3,000,000 PLN (for Maciej Beręsewicz, Marcin Szymkowiak and Kamil Wilak). Authors would like to thank Marcin Augustyniak, Joanna Tyrowicz, Grzegorz Grygiel and Jakub Sawulski for value comments on early version of this paper. All data and codes are available at \url{https://github.com/DepartmentOfStatisticsPUE/rid-gig-economy}.} , 
Nikulin Dagmara\footnote{Gdańsk University of Technology, Poland.}, 
Szymkowiak Marcin\footnote{Poznań University of Economics and Business, Poland; Statistical Office in Poznań, Poland.}, 
Wilak Kamil\footnote{Poznań University of Economics and Business, Poland; Statistical Office in Poznań, Poland.} 
\footnote{The views expressed in this paper are those of the authors and not the Statistics Poland nor Statistical Office in Poznań.}}
\date{}

\begin{document}
\maketitle

\begin{abstract}
In this article we address the question of how to measure the size and characteristics of the platform economy. We propose a~different, to sample surveys, approach based on smartphone data, which are passively collected through programmatic systems as part of online marketing. In~particular, in our study we focus on two types of services: food delivery (Bolt Courier, Takeaway, Glover, Wolt and transport services (Bolt Driver, Free Now, iTaxi and Uber). 


Our results show that the platform economy in Poland is growing. In particular, with respect to food delivery and transportation services performed by means of applications, we observed a~growing trend between January 2018 and December 2020. Taking into account the demographic structure of apps users, our results confirm findings from past studies: the majority of platform workers are young men but the age structure of app users is different for each of the two categories of services. Another surprising finding is that foreigners do not account for the majority of gig workers in Poland. 

When the number of platform workers is compared with corresponding working populations, the estimated share of active app users accounts for about 0.5--2\% of working populations in 9~largest Polish cities. 
\end{abstract}


\clearpage

\section{Introduction}

The changing world of work is characterised by a~growing popularity of labour relationships based on non-standard employment contracts. A~typical indefinite contract for a~full-time job performed on the employer's premises in no longer the~predominant employment model in the contemporary labour market. Digitalisation and flexibilization are contributing to the emergence of new forms of employment. One of the phenomena that has been expanding recently is platform-mediated labour. According to the latest report published by the International Labour Organization (ILO), the number of online web-based and location-based (taxi and delivery) platforms increased from 142 in 2010 to over 777 in 2020 \citep{ILO2021}.

Platform work can be defined as ``non-standard work facilitated by online platforms, which use digital technologies to 'mediate' between individual suppliers (platform workers) and buyers of labour'' \citep[p. 98]{Hauben2020}.  

Originally, the platform economy, also known as \textit{the gig economy} was associated with passenger transport services, mainly with the \textit{Uber} company, which started to arrange work via an online platform. Nowadays, the range of platform-based activities is rapidly growing and includes several types of jobs, such as odd jobs (e.g. \textit{TaskRabbit}), cleaning (e.g. \textit{Helpling}), care (e.g. \textit{care.com}), food delivery (e.g. \textit{Deliveroo}) or programming and translating (e.g. \textit{Upwork}) \citep[cf.][]{Koutsimpogiorgos2020}. This explains why the phenomenon of platform work is receiving more and more attention from researchers and international institutions (cf. e.g. recent reports published by \citet{Eurofound2019} or \citet{ILO2021}). 

Research on these populations is becoming important for several reasons. Primarily, the ``platformisation'' of labour relations is likely to increase competitiveness and create new opportunities on the labour market. Moreover, as employment relations become increasingly less standard, working conditions are likely to deteriorate in terms of job security and legal protections, because platform workers are predominantly self-employed and are not covered by institutional arrangements, such as collective pay agreement schemes. On the other hand, platform work may create opportunities for some groups in the labour market by increasing labour market participation \citep{Eurofound2019} and enabling greater flexibility \citep{Lehdonvirta2018}.

While the number of studies about the platform economy is rapidly increasing, the reported evidence is mostly anecdotal and is based on interviews and personal first-hand experiences \citep{DeStefano2018}. This is because workers in the gig economy, while constantly growing in numbers, constitute a~hard-to-reach and hard-to-identify population \citep[see][Chapter 1]{bohning2017capture}. There is no sampling frame or a~register that provides a~full coverage and members of this group are often indistinguishable from the rest of the population (e.g. they can only be identified by examining the content of their smartphone or laptop). Information about this population is fragmented, which is why it is necessary to resort to modern data sources, such as big data, which can be used for estimating its size and characteristics. Because platform workers tend to operate in urban areas, general population sample surveys do not provide accurate estimates of the reference population (e.g. the economically active population) and administrative data often suffer from over-coverage (e.g. owing to outdated information about the place of residence, delays in reporting, etc.).

In this study, we propose a~new way to measure the size and characteristics of the working age (18--65) population involved in providing food delivery or transportation services, which involves using mobile big data obtained from advertisement systems on smartphones. For this study, we purchased historical data (for 2018--2020) about users of the following apps: \textit{Uber}, \textit{Bolt Driver}, \textit{Glover}, \textit{Wolt Courier}, \textit{TakeAway Courier} and \textit{Bolt Courier}. For comparison, we also purchased data from \textit{iTaxi} and \textit{FREE NOW} apps, which are only used by licensed taxi drivers. Background characteristics, generated by classification algorithms applied by the data provider, include information about location, country of origin, age and sex as well as other characteristics, such as being a~student or having children. In addition, we report descriptive statistics about the average length of app activity on working days (Mondays-Thursdays) and at weekends (Friday-Sunday) during daytime (8--18) and nighttime (18--8).

Our contribution to the literature is twofold. First, we use big data collected via smartphones to passively measure the gig economy in Poland. This is in line with recent trends in official statistics to use all available information rather than create new ones (e.g. surveys). We highlight problems associated with these sources (e.g. measurement error) as well as their strengths (e.g. exact location), which can open a~new avenue for measuring these populations. This approach makes it possible not only to access hard-to-reach and hard-to-identify populations, such as foreigners, but also to plan targeted surveys that can be used to collect more detailed information concerning motivations, working conditions or the quality of life. Second, we provide detailed information regarding the number and background of platform workers, which has not previously been available in Poland at any level of spatial aggregation. We believe that similar data sources can be identified in other countries and therefore our experiences and results may prove useful to other researchers and policy makers. 

The article has the following structure. Section \ref{sec-issues} contains a~review of the literature and provides an overview of the main issues encountered when measuring the gig economy using classical (e.g. sample surveys) and modern data sources (e.g. administrative data, the Internet, big data). Sections \ref{sec-data} and \ref{modern} present data sources about the gig economy in Poland, a~description of selected mobile apps and provide a~coverage assessment of the main data source used in the study. Section \ref{sec-results} presents the main results and compares them with the reference population from the Labour Force Survey (LFS). Finally, Section \ref{sec-conclusions} includes a~discussion of the results and describes future steps in the analysis of the gig economy on the basis of mobile big data.

\section{Issues in measuring the gig economy}\label{sec-issues}

There are a~number of terms to describe the ''platformisation'' of labour relations, including the most common ones, such as ''platform economy'' or ''gig economy'', as well as ''on-demand economy'' or  ''collaborative economy''. The gig economy can be defined in a~broader and narrower sense. The broad definition covers precarious and casual work involving the use of technological intermediation \citep{Aleksynska2019}, while proponents of the~narrower perspective focus either on specific types of platforms or equate the gig economy with digital labour markets, without providing a~detailed definition (for an overview see \citet{Koutsimpogiorgos2020}). Based on the definitions proposed by Eurofound, platform work refers to ``a form of employment that uses an online platform to enable organisations or individuals to access other organisations or individuals to solve problems or to provide services in exchange for payment'' \citep[p. 9]{Eurofound2018}. The authors of the definition also mention some key characteristics of platform work: paid work organised through an online platform; with the involvement of three parties: the online platform, the client and the worker; platform work involves providing specific tasks or solving specific problems; it is outsourced; jobs are broken down into tasks; services are provided on demand \citep{Eurofound2018}. In short, following the review made by \citet{Koutsimpogiorgos2020}, it can be concluded that there is no single definition of the gig economy that is commonly accepted among researchers, policymakers, or practitioners. 

 A~review of the literature shows an~increasing number of studies describing the relatively new phenomenon of the gig economy. What they all have in common is the growing realisation that the long-established idea of a~typical job, which is performed under an indefinite contract, on a~regular basis and during fixed hours, is becoming outdated and is being replaced by new forms of employment. Besides problems with understanding and defining the platform economy, there is a~related concern about how it can be measured. In this section, we provide an overview of past studies aimed at measuring the extent of the platform economy. Please note that we focus solely on the European labour market and do not include results for other countries, which can be found, for example, in \citet{Friedman2014}, \citet{Mitea2018}, \citet{Riggs2019} for New Zealand, \citet{Koustas2019}, \citet{Abraham2018} for US.

Looking at the previous results compiled in Table \ref{tab-estimates-europe} one can see that the majority of past estimates are based on survey data. It is important to underline the numerous limitations related to survey-based estimates mentioned in the literature.  The main concern is about whether surveyed respondents appropriately understand the concept of platform work. As \citet{Bonin2017} point out, respondents may have misclassified other online activities, such using job search websites, search engines or professional networks as forms of platform work.

	 \begin{landscape}
	 \begin{table}[ht]
	     \centering
	     \scriptsize
	     	\caption{Past estimates of the platform economy in Europe}
	     \label{tab-estimates-europe}
	     \begin{tabular}{p{3cm}p{4cm}p{5cm}p{6cm}p{5cm}}
	     \hline
	       Author / Institution   & Data source  & Coverage & the definition of platform work &  Results\\
	     \hline       
	      ETUI Internet and Platform Work Survey \citet{Piasna2019}    &  
	      Face-to-face survey & 
	      Bulgaria, Hungary, Latvia, Poland and Slovakia in 2018--2019; working age adults (aged 18--64); in total, 4,731 respondents & 
	      work using online platforms at least once a~week. Online platforms are defined as websites or apps through which workers can find short jobs or tasks, such as IT work, data entry, delivery, driving, personal services, etc. 
	      &  0.4\% in Poland and Slovakia; 0.5\% in Latvia; 0.8\% in Bulgaria; 1.9\% in Hungary \\
	    \hline       
	    University of Hertfordshire, Foundation for European Progressive Studies, UNI Europa \citet{Huws2019}  & 
	    large-scale online surveys & 
	    11 European countries in 2016--2019 & 
	    platform work conducted at least once a~week & 
	    4.7\% in the UK, 6.2\% in Germany and 9.5\%  in Austria (2016), 28.5\% in the Czech Republic (2019) \\
	    \hline
	   A~Collaborative Economy (COLLEEM) survey conducted by the Joint Research Centre of the European Commission \citet{Brancati2020} & 
	   large-scale online surveys & 
	   2017 and 2018 in 14 and 16 EU member states; 38,878 responses (internet users aged 16--74 years) & 
	   workers who have gained income from providing services via online platforms, where the match between provider and client is made digitally, payment is conducted digitally via the platform, and work is performed either (location-independent) web-based or on-location  & 
	   4.1\% of the adult population in Finland and 9.9\% in the United Kingdom \\
	    \hline
	    ILO 2015 \citet{Berg2016} & 
	    ''1,167 responses of which 814 were from AMT and 353 were from Crowdflower'' &  2015, US and India & Amazon Mechanical Turk and Crowdflower platforms  & Only characteristics of crowd workers  \\
	    \hline
	   
	    \citet{Drahokoupil2019} & 
	    administrative data provided by Smart and a~survey of workers & 
	    Belgium 2016--2018 & 
	    food delivery platforms, Deliveroo  & 
	    3,828 member-riders in October 2017 \\
	    \hline
	    \citet{Aleksynska2019}  & 
	    ''Internet survey; 1000 respondents of the age 18 and older'' & 
	    Ukraine 2017 & 
	    Workers of online platforms are & 
	    Only characteristics of online workers \\
	    \hline
	 
	    \citet{Lee2018}  & 
	    ''self-reported online survey was conducted among Uber users, 295 respondents'' & 
	    Hong Kong 2016 & 
	    Uber users & 
	    analysis of users’ intention of using Uber \\
	    \hline
	     
	     \end{tabular}
	 \end{table}
	 \end{landscape}
	 	 
Moreover, in the case of online surveys, there is a~risk of many inconsistencies caused by the poor reliability of the pooling technique, which may result in the over-representation of those who are more likely to be engaged in platform work \citep{Piasna2019}. Examples of survey-based studies include \citet{Aleksynska2019}, who present survey results for Ukraine; \citet{Piasna2019} presenting estimates for Bulgaria, Hungary, Latvia, Poland and Slovakia; \citet{Berg2016} for the USA and India. Remaining results are based on specific case studies of selected types of platforms, conducted by means of qualitative methods, see among others \citet{Meszmann2018} for Hungary, \citet{Lenaerts} for Belgium, \citet{Sedlakova2018} for Slovakia, \citet{Lee2018} for Hong Kong. However, these case studies provide important knowledge about the characteristics on platform workers rather than estimates of the extent of the phenomenon. Unlike European studies, which are mainly based on survey data, studies conducted in the USA also make use of transaction (financial) data. For example, \citet{Koustas2019} used a~transaction-level dataset obtained from a~large financial aggregator and a~bill-paying application to identify changes in spending patterns for platform workers. Similarly, \citet{Farrell2019} report results based on de-identified checking accounts from Chase bank. To the best of our knowledge, there have been no similar studies based on transactional data that relate to European countries.

\section{Data sources about the gig economy in Poland}\label{sec-data}

\subsection{Probability and non-probability sample surveys}

The main source of information about Internet users in Poland is the Information and Communication Technology (ICT) survey. It has a~standardized methodology, which is used by all EU member states, and provides estimates at country or regional levels based on a~random sample of enterprises and households and its members. Respondents are asked about online activities in the last 3 and 12 months and types of devices they use for connecting to the Internet. For a~detailed description of the design and methodology of the ICT survey see \citet{gus2020ict}. 

While the ICT survey covers the general population, between 2017 and 2020 it investigated the demand side of the platform economy in the question question about transportation services (\textit{Have you used any websites or apps to arrange transportation (e.g. by car) provided by another private person in the last 12 months?}) with the following  options \textit{Yes, dedicated websites or applications (e.g. UBER, BlaBlaCar)}, \textit{yes, other websites or applications (including social networking sites)} and \textit{no}. The proposed answers changed over time and are presented in Appendix \ref{appendix-ict}. In addition, in 2018 and 2019 the survey included two more questions referring directly to the supply side of the gig economy , i.e. \textit{Have you found an assignment through intermediary websites or applications (eg. TakeTask, Sir Local, Upwork, TaskRabbit, Freelancer, Amazon Mechanical Turk) in the last 12 months? (excluding websites of temporary employment agencies / employment agencies )} and \textit{Is the income earned from assignments found via intermediary websites or applications your main or additional source of income?}. Results for these two questions are not reported by Statistics Poland, which may be due to the fact that no or very few respondents answer them positively. The lack of statistics about the supply side can therefore be interpreted to mean that the size of the market for microtasks is negligible in Poland.

The second major survey is the Labour Force Survey (LFS), which measures the size of the active, inactive and unemployed population according to the ILO definition. The LFS is a~quarterly survey with a~rotating panel design; a~detailed description of its methodology can be found in \citet{gus2020bael}. 

At the time of writing the article, the LFS did not include questions or special sections related directly to the gig economy. However, there are plans to include a~special module in the LFS as part of a~coordinated action suggested by the Group of Experts on Measuring Quality of Employment within the United Nations Economic Commission for Europe\footnote{See for instance \url{https://unece.org/statistics/events/group-experts-measuring-quality-employment} and the Handbook on Measuring Quality of Employment (\url{https://unece.org/statistics/publications/handbook-measuring-quality-employment}).}. 

In the article, we used the LFS to obtain estimates of the working 18--65 population at municipality levels in Poland as reference figures for the supply side of the gig economy. Information about the working population at city level is not published by Statistics Poland. Therefore, based on available unit-level data from 2018 to the 1st half of 2020, we obtained estimates and their standard errors according to the Polish LFS methodology. For more details, see Appendix \ref{appendix-lfs}.

Sample surveys conducted by Statistics Poland are not the only source of information about Internet users (further referred to as the online population) or the economically active population. For example, the Office of Electronic Communications\footnote{The President of the Office of Electronic Communications (UKE) is~a regulatory authority responsible for telecommunications and postal activities, as well as the management of frequency resources. It is also an expert body responsible for controlling compliance of products emitting or vulnerable to electromagnetic field emissions, including telecommunication equipment commercially available in Poland. For more details please see \url{https://www.uke.gov.pl/en/about-us/}.} publishes an annual report on the functioning of the telecommunications services market and consumer preferences, which is based on a~survey\footnote{Key findings about individual customers can be found at \url{ https://uke.gov.pl/en/newsroom/consumer-survey-2020-individual-customers,333.html}} conducted by commercial market research companies.. The survey covers a~variety of topics, such as mobile phone customers, Internet access or the use of digital media. It provides more detailed information about the online population but the sample size is significantly smaller than in surveys conducted by Statistics Poland (about 1600-2000) and a~detailed description of the methodology is not published. We used this survey as a~source of estimates of the level of smartphone coverage in Poland.

Another source that we considered is a~non-probability sample of Internet users conducted by the PBI company (the acronym stands for \textit{Polskie Badania Internetu}, which means `Polish Internet research'). PBI specialises in measuring online audiences and their characteristics. Between 2016 and 2020 PBI conducted the Megapanel PBI/Gemius survey, which is exclusively focused on Internet users. Participants were were asked to install a~special computer application that tracked all their online activity. About 150,000 respondents were recruited to participate in the survey by pop-ups placed on major Polish websites. 

In 2020, PBI, together with the Gemius company\footnote{Gemius is an international research and technology company providing data, as well as advanced tools for digital and traditional marketing activities such as web analytics, online campaigns’ management and ad serving. For more see: \url{https://www.gemius.com/about-us.html}.} and the Radio Research Committee\footnote{More information can be found at \url{https://badaniaradiowe.pl/}.} launched a~new survey called the Mediapanel. It is a~combination of the Internet research standard, the radio audience standard (Radio Track) and data from the single-source cross-media Gemius measurement. The Mediapanel includes a~passive measurement of Internet, TV and radio consumption. In 2021 the panel had over 280,000 users.

We asked PBI to provide an estimate of the number of users between 2018 and 2020 using at least one of 25 websites devoted to freelancing or app testing, such as \url{www.mturk.com}, \url{www.clickworker.com}, \url{www.fiverr.com} or \url{www.upwork.com}\footnote{For the whole list please consult Appendix \ref{appen-websites}.}. As in the previous cases, almost none of the above-mentioned websites met the requirement of the minimum number of active users to provide reliable estimates.


It can therefore be concluded that the existing survey-based data are not sufficient to accurately measure the size of the gig economy in Poland. Moreover, given their limited sample sizes, these surveys can at best provide reliable estimates at the national level, while the gig economy is connected with the use of the Internet and smartphones at municipal level. There are no sources that provide reliable estimates at low levels of spatial aggregation or for detailed cross-classifications. One way to overcome this problem is to turn to administrative data or Internet data sources (big data).

\subsection{Administrative data}


Passenger transport in Poland is regulated by the provisions of the Road Transport Act (Pol. \textit{Ustawa z~dnia 6 września 2001 r. o~transporcie drogowym, Dz.U. 2001 nr 125 poz. 1371}). According to the act, providers transport services, including taxi drivers, are obliged to meet certain requirements, such as having an appropriate license. Until 2019, however, the act did not apply to private persons providing transportation services. This legal loophole enabled applications transport companies like Uber to operate in Poland, which practically meant that any person meeting the requirements imposed by their applications but not necessarily meeting the legal requirements for taxi drivers, could provide transport services. In the case of Uber, however, there was another problem associated with the fact that payments for Uber services were made without the use of a~cash register. That practice was out of line with the regulation on cash registers that came into effect in 2002 (\textit{Rozporządzenie Ministra Finansów z~dnia 20 grudnia 2001 r. w~sprawie kas rejestrujących, Dz.U. 2001 nr 151 poz. 1711}), by virtue of which all entities providing services in transport, including private individuals, have been obliged to register their turnover by using a~cash register. In practice this means that vehicles used for transporting paying customers have to be fitted with a~cash register, which was not the case with Uber drivers.

The lack of coherent legislation on road transport has had negative consequences. The regulation that was did not take into account the new forms of services sparked off conflicts between taxi and Uber drivers. Faced with higher operating costs and increased competition, taxi drivers felt a~growing sense of injustice, On the other hand, Uber drivers were subjected to checks and had to pay financial penalties for not having cash registers.

In view of the growing popularity of applications such as Uber, the situation was remedied  by an amendment to the Road Transport Act (\textit{Ustawa z dnia 16 maja 2019 r. o zmianie ustawy o transporcie drogowym oraz niektórych innych ustaw, Dz.U. 2019 poz. 1180}), which came to be known as ``Uber lex''. All the changes intriduced in the new act now apply to entrepreneurs and private individuals providing transport services to ensure that taxi and Uber drivers play by the same rules. The Uber lex extended the license obligation to Uber drivers, but at the same time simplified the licensing procedure. The problem of recording turnover has been solved by enabling the use of virtual cash registers. The Uber lex came into effect in 2020 with a~transition period until the end of March, which was then extended to the end of September and, later, to the end of December owing to the pandemic.

 A~taxi license is valid within the municipality in which it was issued by the head of the municipal government. Databases of licensed drivers are created independently by administrators and there is no common IT system that would enable the creation of a~central register. Moreover, in practice, data is often collected in paper form. According to the act, records of persons with a~taxi license do not have to include information about the method of accounting  used by drivers for recording turnover (cash register / virtual cash register). As a~result, it is impossible to identify drivers who provide transport services by means of applications. 

Taking into account the above-mentioned limitations concerning records of taxi license holders, and given the fact that the license obligation for Uber drivers was imposed only in 2020, with a~transition period lasting until the end of the year, data from municipal registers are not suitable for the purpose of measuring the gig economy in Poland.

\section{Big data from smartphones}\label{modern}

\subsection{Programmatic advertisement system}

According to Garner IT Glossary, big data sources are described as: ``high-volume, high-velocity and/or high-variety information assets that demand cost-effective, innovative forms of information processing that enable enhanced insight, decision making, and process automation''. In addition to generating new commercial opportunities in the private sector, big data are potentially a~very interesting data source for official statistics, either to be used on their own, or in combination with more traditional data sources, such as sample surveys and administrative registers \citep[cf.][]{daas2015big}. 

Big data sources are used for studying the gig economy from different angles. For instance, change in income at the~national level has been measured based on transaction-level datasets, such as those generated by bill-paying applications \citep[cf.][]{Koustas2019}; detailed behaviour and matching between customers and workers have been analysed using detailed data generated by Uber\footnote{See: \url{https://eng.uber.com/research/}.} or Lyft\footnote{See \url{https://eng.lyft.com/tagged/data-science}.}. However, access to such data is restricted and often granted only to researchers from financial institutions or providers of the apps.

In this study, we also used big data but from a~different source. We purchased aggregated data from the Selectivv company\footnote{For more information see \url{https://selectivv.com/en/}.}, which uses a~\textit{programmatic} advertisement platform to monitor about 350,000 mobile apps used and over 17 million mobile websites visited by over 20 million smartphone users in Poland. 

\textit{Programmatic advertising platforms} are IT systems that automate the process of displaying advertisements to Internet users based on their individual characteristics, such as past online behavior patterns or location. It is based on real-time bidding systems that in a~matter of milliseconds decide whether to show a~given ad to a~given user \citep{busch2016programmatic}. 

In practice, a~programmatic system works as follows. When a~mobile device user opens a~website or an~application that contains ad space with an advertisement displayed by a~particular company (e.g. Selectivv), the company receives basic information from the mobile device along with an individual number associated with the user account. This number is assigned through two systems: \textit{the Google advertising ID} (GAID) and \textit{Apple's Identifier for Advertisers} (IDFA). The first one works on smartphones with Android, the second one -- on those with iOS. The number remains the same as long as the user logs in with their Google/iOS or another account. If the user changes a~mobile phone and creates a~new account, then they are treated as a~new user. However, based on communications with Selectivv's staff, this situation is very rare.

The main difference between data collected through programmatic platforms and systems of mobile network operators is the level of available information. The latter can only provide information about customers, often without a~distinction between business and private users, about location, based on signalling, or the type of smartphone. Most of these data are taken from Call Detail Records (CDR), which, according to Polish law, are separated from consumer relationship management systems. Thus, data collected by mobile network operators contain no information about apps installed on a~give phone or any background details about a~given user.

On the other hand, companies working with programmatic systems serve multiple providers and collect information not only about domestic users but also about foreign ones, who are often missed by surveys or registers. As almost all apps contain advertisements, the amount of information collected is considerable and may include: location (GPS / WiFi), time of use, websites visited, applications installed, device type, operating system and its settings (such as language) and mobile network operator. Programmatic systems do not have access to activities within mobile applications and only provide information about the app name, time spent in the app as well as when and where it was used (depending on smartphone settings). 

Based on passively collected data, companies profile users of mobile devices by applying machine learning algorithms or heuristic rules. In our case, Selectivv claims that their algorithms predict with high accuracy up to 360 variables about a~single user. While the store of collected information is rich, it does not contain any personal details, such as name, surname or personal ID, except for user characteristics. Selectivv also employs algorithms to identify users of multiple devices (private and business ones) based on GPS location and WiFi signals.

In the next subsections we will describe variables used in this study, assess the coverage of Selectivv data by comparing them with existing official statistics and administrative data and describe mobile apps selected with the aim of measuring the size of a~gig economy.

\subsection{Variables obtained from smartphones}

We obtained aggregated data for various socio-demographic characteristics describing smartphones users, which had been specified during a~discussion with Selectivv's staff. The type and levels of all variables were limited by the project budget. All variables are the output of Selectivv's proprietary classification or heuristic algorithms, so the level of errors associated with these methods is unknown. For the study, we selected the following variables:

\begin{itemize}
    \item \texttt{Sex} (Male, Female) -- the classification was made on the basis of a~user's activity (e.g. visited websites, installed applications, information provided in the apps) and sample surveys conducted by Selectivv via advertisement systems.
    \item \texttt{Age group} (18--30, 31--50, 51--64) -- the classification was made on the basis of a~user's activity (e.g. visited websites, installed applications, information provided in the apps) and sample surveys by Selectivv via advertisement systems.
    \item \texttt{Nationality} (Polish, Ukrainian, Other) -- Ukrainians are defined as people who have a~SIM card provided by a~Polish operator, have set the language on their smartphone to Russian or Ukrainian and at least once in the last year have been to Ukraine, where they replaced their SIM card with another one provided by a~Ukrainian operator.
    \item \texttt{Residence} (Cities, Functional Urban Areas and Province; for more see Appendix \ref{appen-fua}) -- this information is derived on the basis of location metadata (e.g. GPS, WiFi) and is defined as the most frequent night location (18:00--8:00) in a~given period. We provided a~shapefile with borders of these areas.
    \item \texttt{Student} -- whether a~given person is a~student; derived based on a~user's location data (e.g. points of interest; POI) and their browsing history.
    \item \texttt{Parent of child aged 0--4} -- derived on the basis of POIs, such as kindergartens, visits at offline and online shops.
    \item \texttt{Parent of child aged 5--8} -- derived on the basis of POIs, such as primary schools, visits at offline and online shops.
    \item \texttt{Time spent in the app} -- the mean and standard deviation of the time spent in the app (in seconds).
\end{itemize}

For the purpose of our study, we defined an active user as \textit{a~person who used a~given app for at least one minute} within a~given period (month, half-year between 2018 and 2020). We decided not to use thresholds such as one, two or more hours because prior to the analysis we did not know how long particular users use selected apps. Moreover, as these apps are designed for drivers and couriers they are more likely to use them frequently. Finally, the pragmatic reason was our~limited budget, which was only sufficient to obtain data for long periods, such as months or half-years. 

We obtained four datasets for a~period from 2018 to 2020 for the population of users aged 18--64:

\begin{enumerate}
    \item the number of active users by app within a~~given month,
    \item the number of active users by sex, age group and nationality within a~given half-year; for each cross-classification we got a~share of students, parents with children aged 0--4 and parents with children aged 0--4,
    \item the number of active users by city, functional urban area and province within a~given half-year,
    \item the mean and standard deviation of the time spent in the app for the following periods: Mondays-Thursdays 8:00-18:00, Mondays-Thursdays 18:00--8:00; Fridays-Sundays 8:00-18:00 and Fridays-Sundays 18:00--8:00 within a~given half-year.
\end{enumerate}

All datasets used in the study are available as supplementary materials included with this article.

\subsection{Assessment of the coverage error} 

According to the company's website, in 2019 Selectivv collected data about 21 million unique smartphone users in Poland\footnote{See \url{https://index.selectivv.com/}.}. This number includes Polish and non-Polish users. According to consumer satisfaction surveys conducted by the Office of Electronic Communications (OEC), 22.5 million Polish citizens \footnote{See \url{https://uke.gov.pl/akt/badania-konsumenckie-2019,286.html} as of November 2019, 70\% of people aged 16+ had a~smartphone (96\% owned a~private device and 6\% had a~business one). Note that the survey involved on a~sample of 1600 respondents, so the results are likely to be uncertain.} had a~smartphone. This suggest that the coverage of Selectivv data is high. Unfortunately, the company does not report any background information regarding the spatial distribution of the data, which limits comparability. 

One way to assess coverage is to verify the share and changes in the population of foreigners in Selectivv databases. This population is often missed by official statistics owing to the lack of up-to-date sampling frames. In a~recent report \citep{selectiv2021}, the number of Ukrainians reported in January 2021 was 1.274 million.  According to recent experimental studies conducted by  \citet{gus2020appendix,gus2020}, there were 1.351 million Ukrainians on 31st December 2019. The latter number was obtained by integrating 9 administrative sources concerning the registered population, in particular the economically active population. It can therefore be concluded that Selectivv data provide an adequate coverage of the Ukrainian population.

\begin{figure}[ht!]
    \centering
    \includegraphics[width=0.8\textwidth]{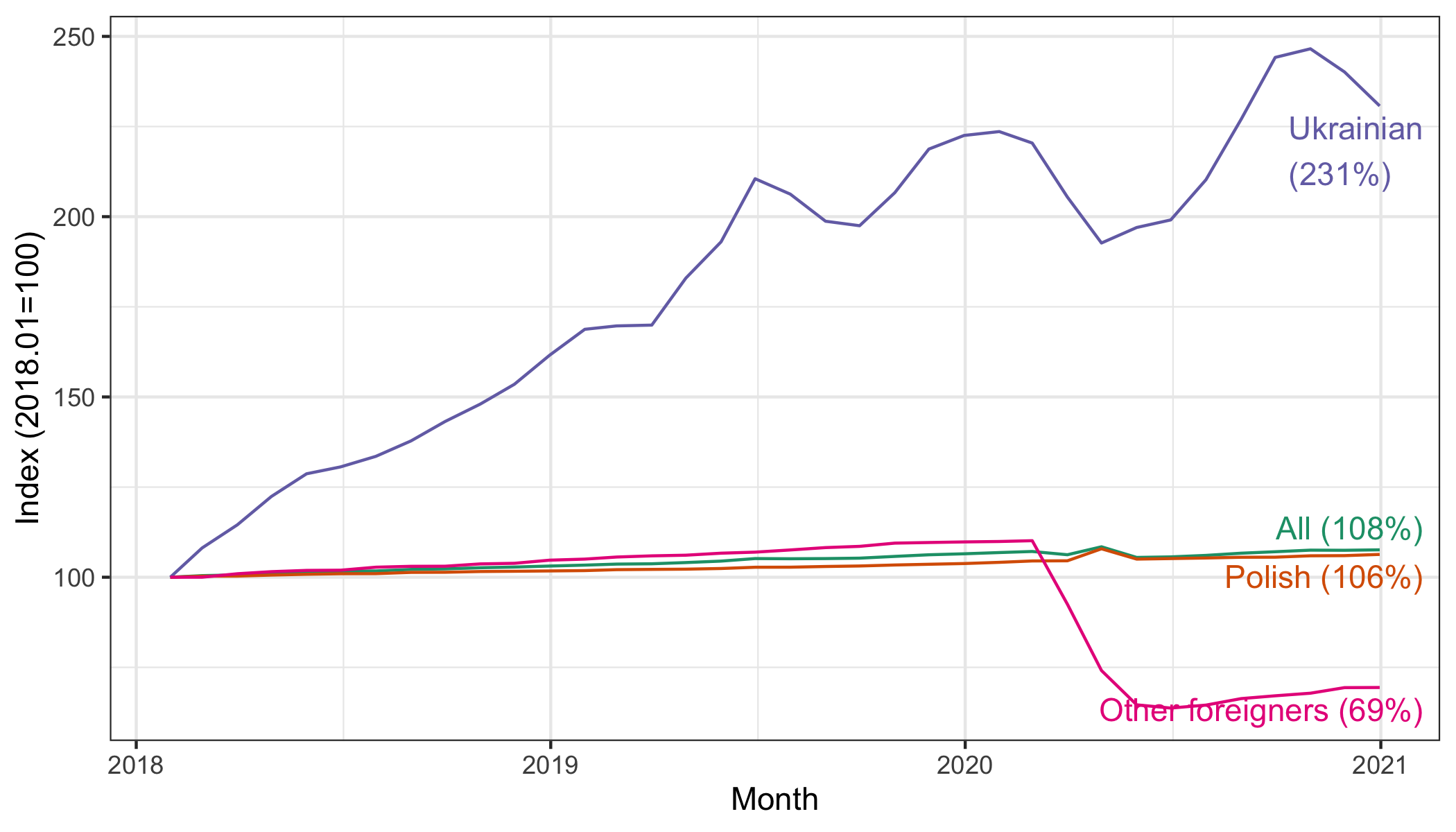}
    \caption{Changes in the number of smartphone users based on monthly data collected by Selectivv between 2018 and 2020 (2018.01 = 100)}
    \label{fig-sel-index-total}
\end{figure}

Figure \ref{fig-sel-index-total} presents changes in the population of Selectivv users by country of origin. Between 2018 and 2020 the number of all smartphone users in their database increased by 8\% but the change for each sub-population was different. The group of Polish users increased by 6\%, which, in absolute numbers, corresponds to about 2 million additional users between 2018 and 2020. 

The biggest changes can be observed for Ukrainians and other foreigners, whose number fell considerably after the first COVID lockdown in March 2020, but, unlike in the case of Ukrainians, it did not change much over the rest of 2020.  In contrast, the number of Ukrainian users rose by over 200\% between 2018 and 2020, which is the result of migration to Poland and can also be observed in official sources.

To further verify the degree of coverage of the foreign population, we compared these changes with the index of insured foreigners, calculated on the basis of records maintained by the Social Insurance Institution (ZUS). If a~foreigner is legally employed in Poland, they should be listed in the ZUS register. The register includes economically active foreigners and has been consistent with the structure of migration to Poland since 2018. 

\begin{figure}[ht!]
    \centering
    \includegraphics[width=
    0.9\textwidth]{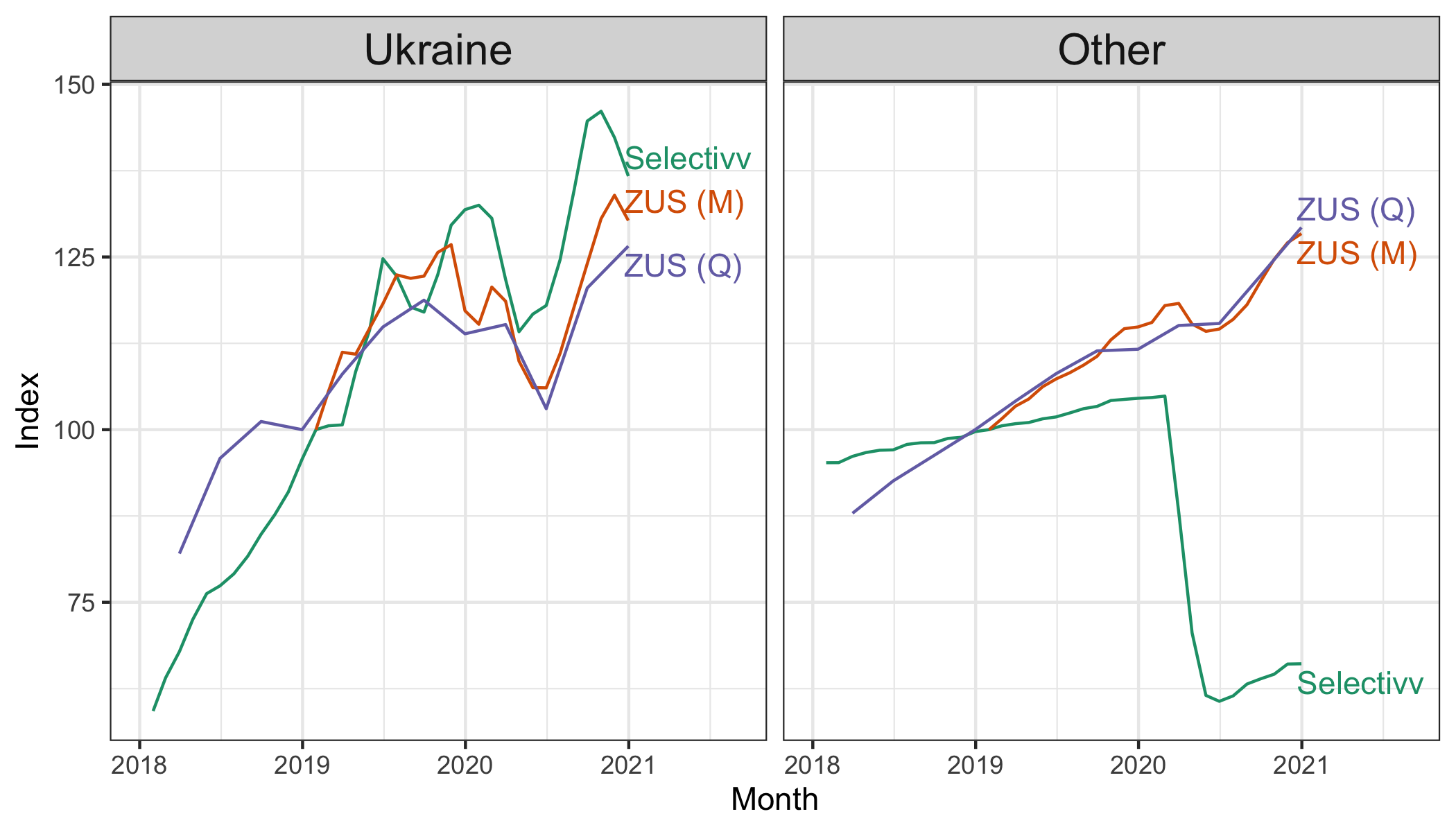}
    \caption{ A~comparison of changes in the number of users according to Selectivv and the number of insured foreigners according to ZUS (2019.01 = 100). Note: ZUS (M) -- monthly statistics; ZUS (Q) -- quarterly statistics}
    \label{fig-sel-zus}
\end{figure}

Figure \ref{fig-sel-zus} presents a~comparison between monthly and quarterly data from ZUS and Selectivv data for two groups of foreigners: Ukrainians and non-Ukrainians (other). Trend lines for Ukrainians, which are evident in ZUS data, in particular the sharp drop and increase in 2020, are similar to those observed in Selectivv data. The main difference between Selectivv and ZUS regarding this subgroup are the actual index values (January 2019 is used as the~baseline). It should be remembered, however, that not all Ukrainians would have applied for a~legal job and that Selectivv also covers children. 

In contrast, the drop in the index for non-Ukrainian foreigners is significantly bigger than that observed in ZUS registers. There are three possible explanations for this. First, since we do not know the number of other foreigners in Selectivv databases, the decrease may simply be due to their low initial levels. In ZUS data there are about 200,000 non-Ukrainian foreigners. Second, ZUS covers all foreigners, including those who may have lived in Poland for several years and are already assimilated (e.g. use the Polish language), while Selectivv may be covering short-term migrant workers. Finally, the drop may represent illegal workers who lost their source of income during the COVID pandemic.

After comparing Selectivv data with existing official or administrative sources, it can be concluded with a~high degree of confidence that our data source covers the majority of smartphone users in Poland. Certainly, access to more detailed information could improve our assessment but as far as we know there is no other available data source in Poland with this level of coverage. Therefore we are convinced that these data can provide accurate information regarding the number of  people engaged in the gig economy in Poland. 

\subsection{Applications used in the study}

In the following step we examined mobile apps that could be used to identify gig workers. Table \ref{tab-apps} contains a~list of such applications, divided into the following categories: transportation, delivery, services, crowdwork and microtasks.  In each case, we identified the provider, i.e. the company that develops a~given app, the name of the app for workers and additional information. 

\clearpage

\begin{table}[ht!]
\centering
\caption{Mobile applications by category, with an indication of the target group of users}
\label{tab-apps}
\small
\begin{tabular}{llll}
  \hline
Category & Provider & The name of the app for workers & Comments \\ 
  \hline
  Transportation & Uber & Uber Driver &  one app for Uber;\checkmark \\ 
   & Bolt & Bolt Driver &  \checkmark \\ 
   & Lyft & Lyft Driver &  no or small \# users  \\ 
   & FREENOW (Mytaxi) & FREE NOW for drivers &  \checkmark\\ 
   & iTaxi & iTaxi Kierowca K3 & \checkmark \\ 
   & Optitaxi &  -- & no or small \# users \\ 
  \hline
  Delivery 
  & Deliveroo & Deliveroo Rider &  no or small \# users \\ 
  & Glovo & Glover & \checkmark \\ 
  & UberEats & Uber Driver & one app for Uber; \checkmark \\ 
  & Coopcycle &  -- & GP 5k+ \\ 
  & Take Eat Easy &  -- &  no or small \# users  \\ 
  & Pyszne.pl & Takeaway.com Courier & Same as pyszne.pl; \checkmark \\ 
  & Lieferando & Takeaway.com Courier & Same as pyszne.pl; \checkmark \\ 
  & Takeaway & Takeaway.com Courier & Same as pyszne.pl; \checkmark \\ 
  & Bolt Food & Bolt Courier &  \checkmark\\ 
  & Wolt & Wolt Courier Partner & \checkmark \\ 
  \hline
  Services
  & TaskRabbit & Tasker by TaskRabbit  & GP 100k+ \\ 
  & Helpling & Helpling Partner & no or small \# users  \\ 
  & Fiverr &  -- & GP 10 mln+ \\ 
  & Upwork & Upwork for Freelancers &  no or small \# users \\ 
  & Freelancer & -- & GP 5 mln+ \\ 
  & PeoplePerHour & -- & GP 100k+ \\ 
  & Toptal &  -- & GP 10k+ \\ 
  & Guru &   -- & GP 5k+ \\ 
  & FlexJobs &   -- & GP 1k+ \\ 
  & Truelancer &   -- & GP 500k+ \\ 
  & ClearVoice &  -- & -- \\ 
  \hline
  Crowdwork 
  & Amazon Mechanical Turk & Turkdroid & GP 50k+ \\ 
  & Clickworker & -- & GP 500k+ \\ 
  & Microworkers & -- & GP 5k+ \\ 
  & CrowdFlower & -- & -- \\ 
  & Spare5 & -- & -- \\ 
  \hline
  Microtasks 
  & App Jobber & -- & GP 100k+ \\ 
  & ShopScout &  -- & GP 50k+ \\ 
  & Streetspotr & -- & GP 500k+ \\ 
   \hline
\end{tabular}
\begin{flushleft}
Note: Symbol '--' in the column \textit{The name of the app for workers} denotes that there were no separate app for workers (i.e. only one app for clients and workers); GP -- number of downloads from Google Play; \checkmark -- app was selected for the study.
\end{flushleft}
\end{table}

\clearpage

This is a~crucial part of our analysis. We decided to focus on dedicated apps for workers in order to distinguish between the demand and supply side. We found that the majority of transportation and delivery platforms create apps for workers, while platforms for services, crowdwork or microtasks tend to have only one app for both workers and customers.  It is important to point out that Uber uses only one app to provide transportation (Uber) and delivery services (Uber Eats). Programmatic services do not have access to user activity within the app, so we only have information about the time spent in the app. 

The \textit{Comments} column in Table \ref{tab-apps} specifies whether it was possible to use a~given app in the study. For instance, three providers (Pyszne.pl, lieferando and takeaway) use the same app: Takeaway.com Courier. Moreover, information about the number of downloads shown in the Google Play store or the App store was misleading. For example, while TaskRabbit has been downloaded over 100,000 times and Fiverr -- over 10 million times, data provided by Selectiv indicated that no user in their databases had installed either of these two apps. This suggests that Google Play probably presents information about the global number of downloads, not just for Poland.

Finally, taking into account these limitations and our main assumptions, we selected 8 apps for further analysis: \textit{Uber}, \textit{Bolt Driver}, \textit{Glover}, \textit{Wolt Courier}, \textit{Takeaway.com Courier} and \textit{Bolt Courier}. For the purpose of comparison, we used \textit{iTaxi} and \textit{FREE NOW}, which are designed for licensed taxi drivers. A detailed description of the apps is provided in Appendix \ref{appendix-apps}.

\section{Results}\label{sec-results}

\begin{figure}[ht!]
    \centering
    \includegraphics[width=0.8\textwidth]{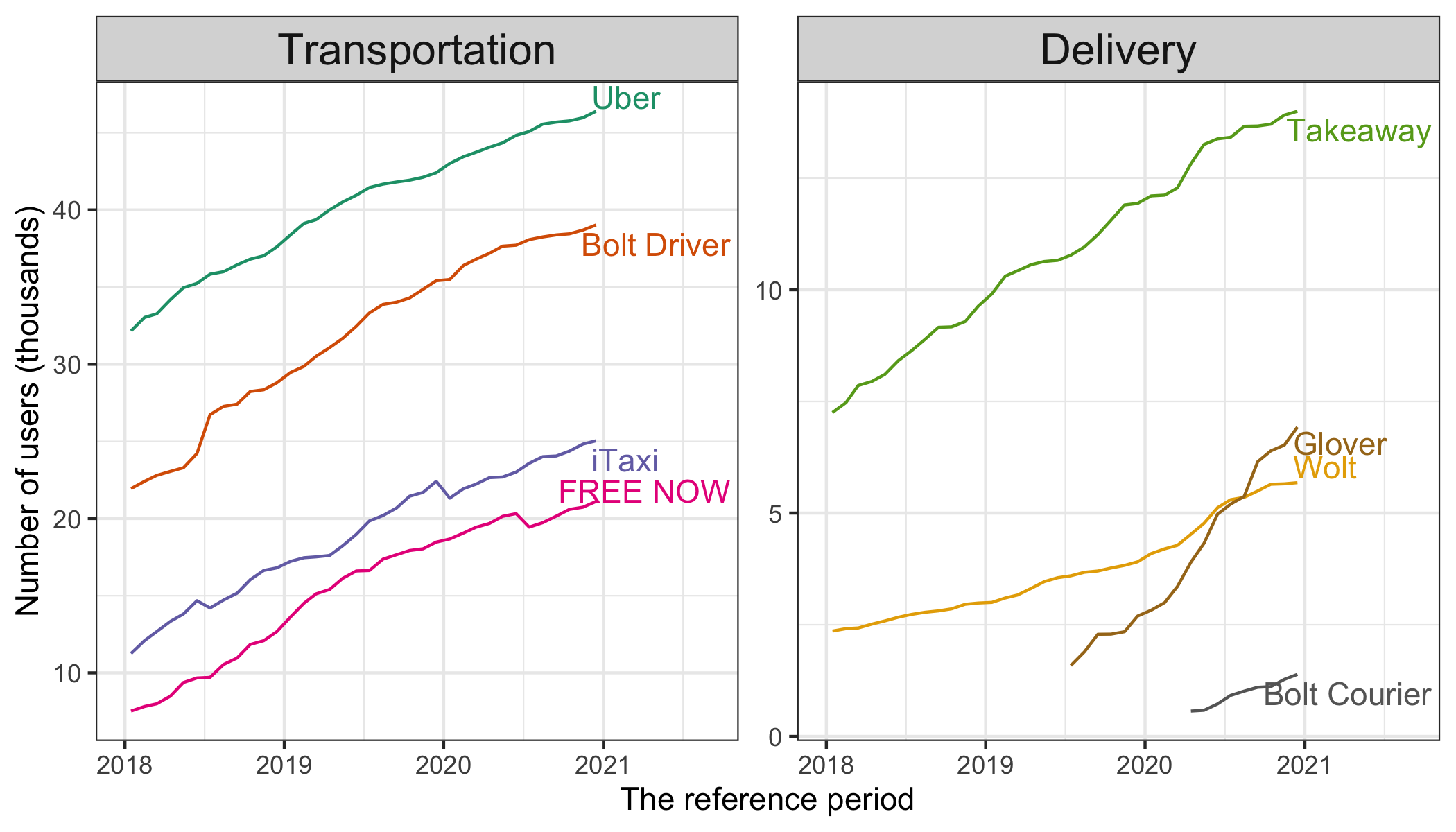}
    \caption{The monthly number of active users in Poland for selected apps by category between 2018 and 2020}
    \label{fig-month}
\end{figure}

Figure \ref{fig-month} presents changes in the number of active users (in thousands) between January 2018 and December 2020. Data for Glover and Bolt Courier are only available for a~shorter period because these services started operating in Poland in mid 2019 and beginning of 2020. Interestingly, the rate of growth for Glover is significantly higher than for the other apps. This is mainly due to the fact the company offers delivery services not just for meals, including fast-food meals, such as those sold by McDonald's, but also groceries from the biggest chain store in Poland -- Biedronka. 

In the case of other delivery apps, Takeaway saw the highest absolute increase, from 7,500 to nearly 14,000 users in December 2020. On the other hand, at the end of 2020, Wolt and Bolt Couriers had around 6,000 and 2,000 users, respectively. These disparities have to do with the companies' regional scope of operation. For instance, only 14\% of Takeaway users were from Warsaw (Poland's capital city) while the corresponding share for the other apps was between 35\% and 40\%. 

\begin{figure}[ht!]
    \centering
    \includegraphics[width=0.8\textwidth]{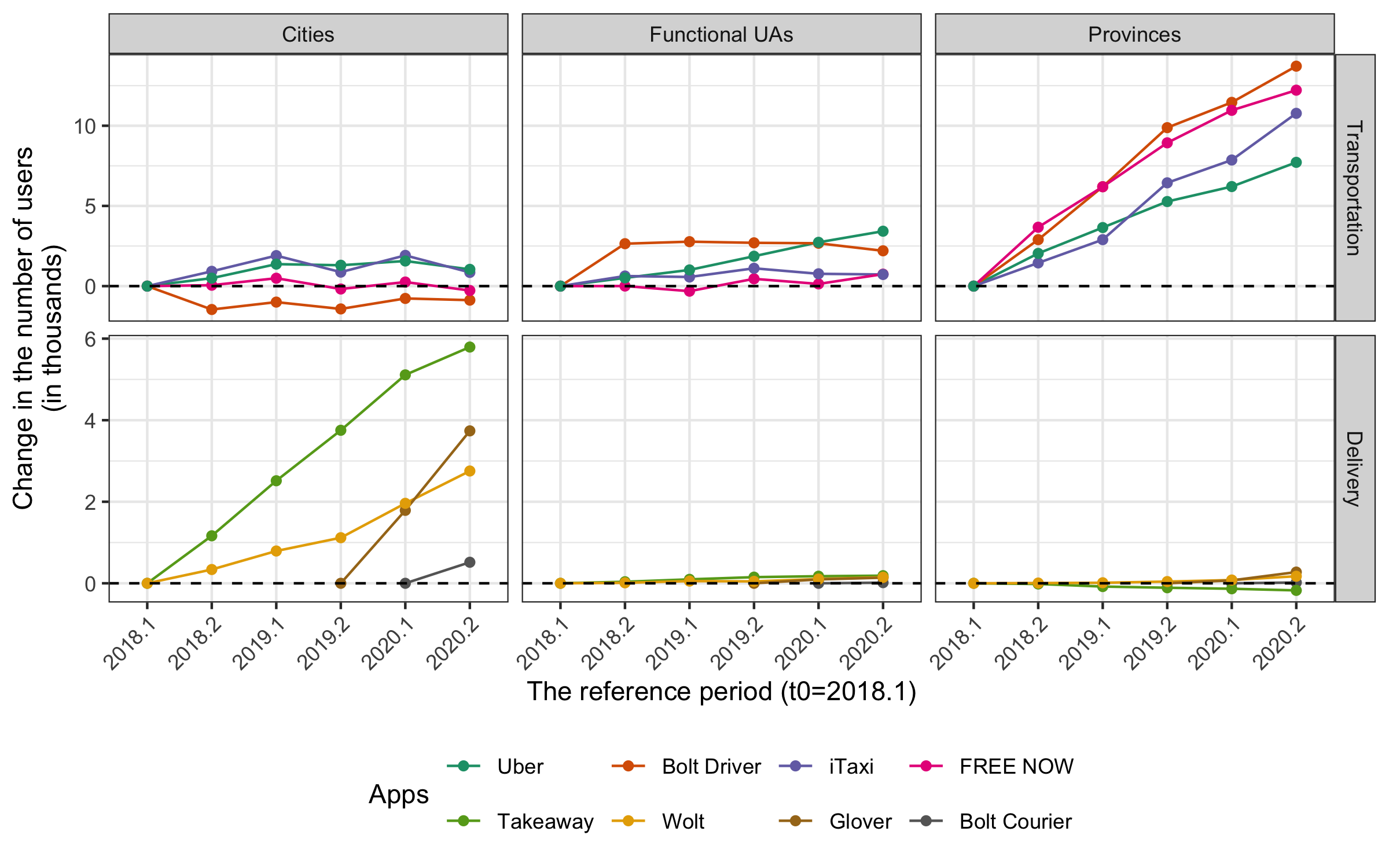}
    \caption{Change in the number of active users in Poland for selected apps by type between 2018 and 2020. Base 2018HY1}
    \label{fig-diff-hy}
\end{figure}

Figure \ref{fig-diff-hy} indicates where the increases presented in Figure \ref{fig-month} actually happened. In the case of delivery apps, the biggest increases were recorded in cities, which is to be expected given that these services are mainly offered in cities. Changes in functional urban areas (FUAs) and provinces are negligible relative to 2018HY1. In the case of transportation services, changes in cities and FUAs are small and there is even a~decrease in the number of Bolt Driver users. The observed increase in the number of users is mainly due to the expansion of these services to areas surrounding cities, which is evident at the province level. For instance, in the case of Tricity (the metropolitan area of Gdansk, Sopot and Gdynia) services offered by Uber actually cover almost the entire Pomorskie province (as of May 2021).

For more details, see Appendix \ref{appen-city-levels}.  With regard to these results, it should be noted that the same users may have been using more than one of these apps, but the level of this overlap is unknown.

Table \ref{tab-demo-table1} contains demographic characteristics of active apps users for the last period in our analysis (2020HY2), which may be indicative of the self-selection process. The main difference between transportation and delivery services is the age structure. The majority (95\%) of couriers were between 18 and 30, which is mainly related to the mode of transportation -- bikes and scooters. There is also a~major difference in terms of sex in both categories, with female drivers accounting for only 11-12\% of all users, and an even smaller share of female couriers (5-10\%). This finding is in line with survey results showing that, on average, platform work is mainly performed by younger people \citep{Piasna2019, Brancati2020}.

Smaller shares of foreigners of other nationalities, students and parents with children aged 0-4 are, most likely, to do with age. The majority of recent migrants in Poland are young people from Ukraine, Belarus, Russia, India, Bangladesh or Nepal. 

\begin{table}[ht!]
\centering
\small
\caption{Demographic characteristics of gig app users in 2020HY2}
\label{tab-demo-table1}
\begin{tabular}{lrrrrrrrrrrr}
  \hline
  & \multicolumn{2}{c}{Sex} & \multicolumn{3}{c}{Age group} & \multicolumn{3}{c}{Nationality} & Student  & \multicolumn{2}{c}{Parent of a~child} \\
  \cline{2-12}
App & Men & Women & 18--30 & 31--50 & 51--64 & PL & UA & other &  & aged 0--4 & aged 5--10 \\ 
   \hline
  \multicolumn{12}{c}{Transportation} \\ 
  \hline
  Uber  & 88.0 & 12.0 & 49.2 & 46.4 & 4.4  & 66.0 & 24.1 & 9.8 & 2.7 & 3.5 & 3.1\\ 
  Bolt  & 86.3 & 13.7 & 56.2 & 37.8 & 6.0  & 65.5 & 26.1 & 8.5 & 3.1 & 4.4 & 1.9  \\ 
  FREE NOW  & 88.6 & 11.4 & 40.1 & 52.4 & 7.5  & 76.8 & 21.7 & 1.5 & 0.8 & 2.0 & 3.2  \\ 
  iTaxi  & 88.7 & 11.3 & {28.2} & {58.1} & {13.7}  & 77.9 & 20.2 & 2.0 & 4.4 & 0.6 & 1.1 \\ 
    \hline
  \multicolumn{12}{c}{Delivery} \\ 
  \hline

  Takeaway & 89.7 & 10.3 & {94.6} & 4.5 & 0.9 & 62.1 & 31.1 & 6.7 & 8.1 & 8.1 & 1.6  \\ 
  Glover & 93.8 & 6.2 & {94.1} & 5.7 & 0.2  & 61.6 & 27.5 & 10.8 & 7.8 & 7.8 & 1.4  \\ 
  Wolt  & 92.3 & 7.7 & {95.7} & 2.9 & 1.4  & 54.2 & 28.4 & 17.5 & 7.9 & 8.0 & 1.5  \\
  Bolt Courier & 94.7 & 5.3 & {100.0} &  -- & --  & 62.3 & 27.5 & 10.2 & 0.0 & 0.0 & 0.0  \\
 \hline
\end{tabular}
\begin{flushleft}
Note: PL -- Polish, UA -- Ukrainian, other -- other foreigners.
\end{flushleft}
\end{table}

\begin{figure}[ht!]
    \centering
    \includegraphics[width=0.9\textwidth]{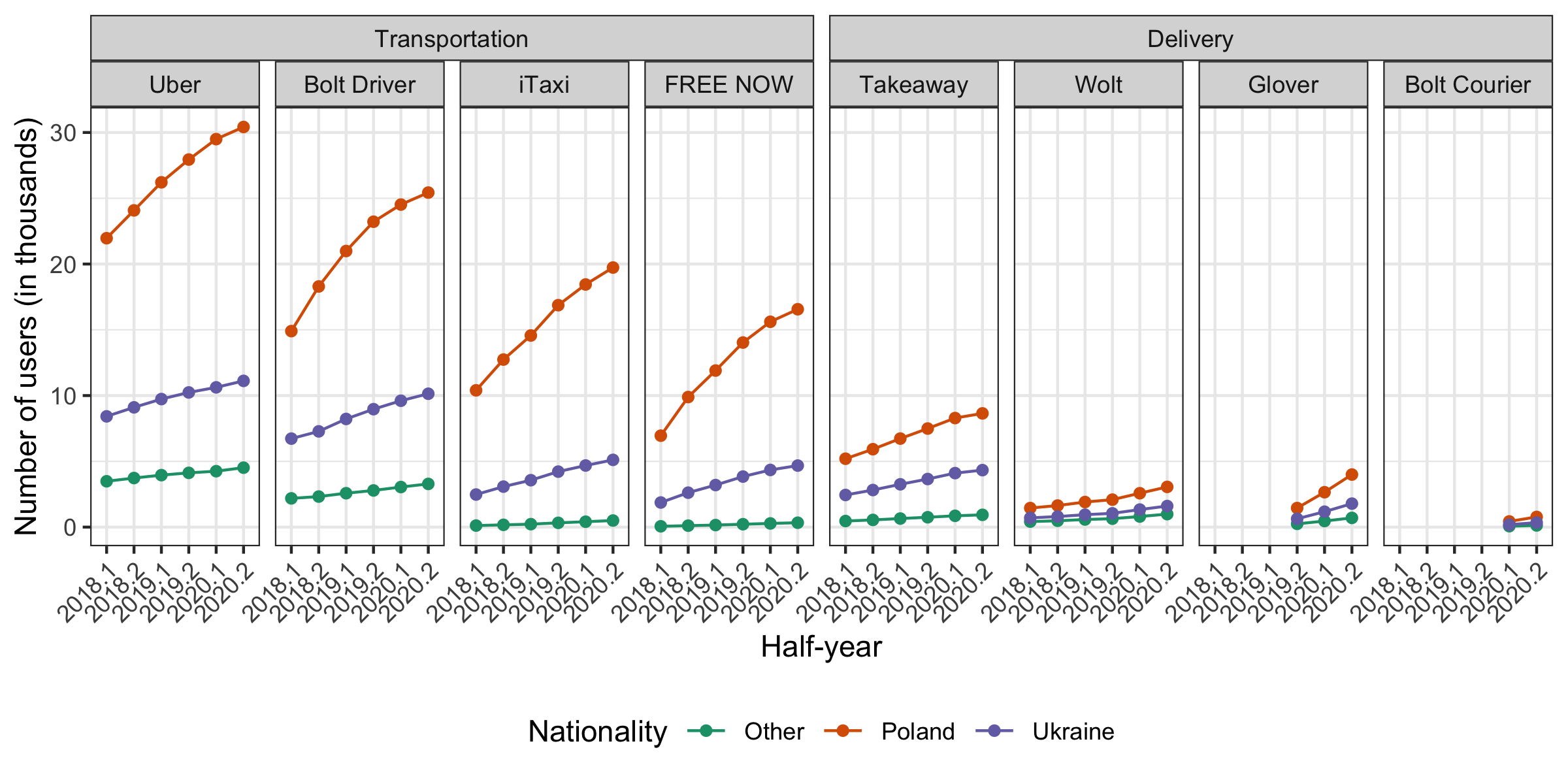}
    \includegraphics[width=0.9\textwidth]{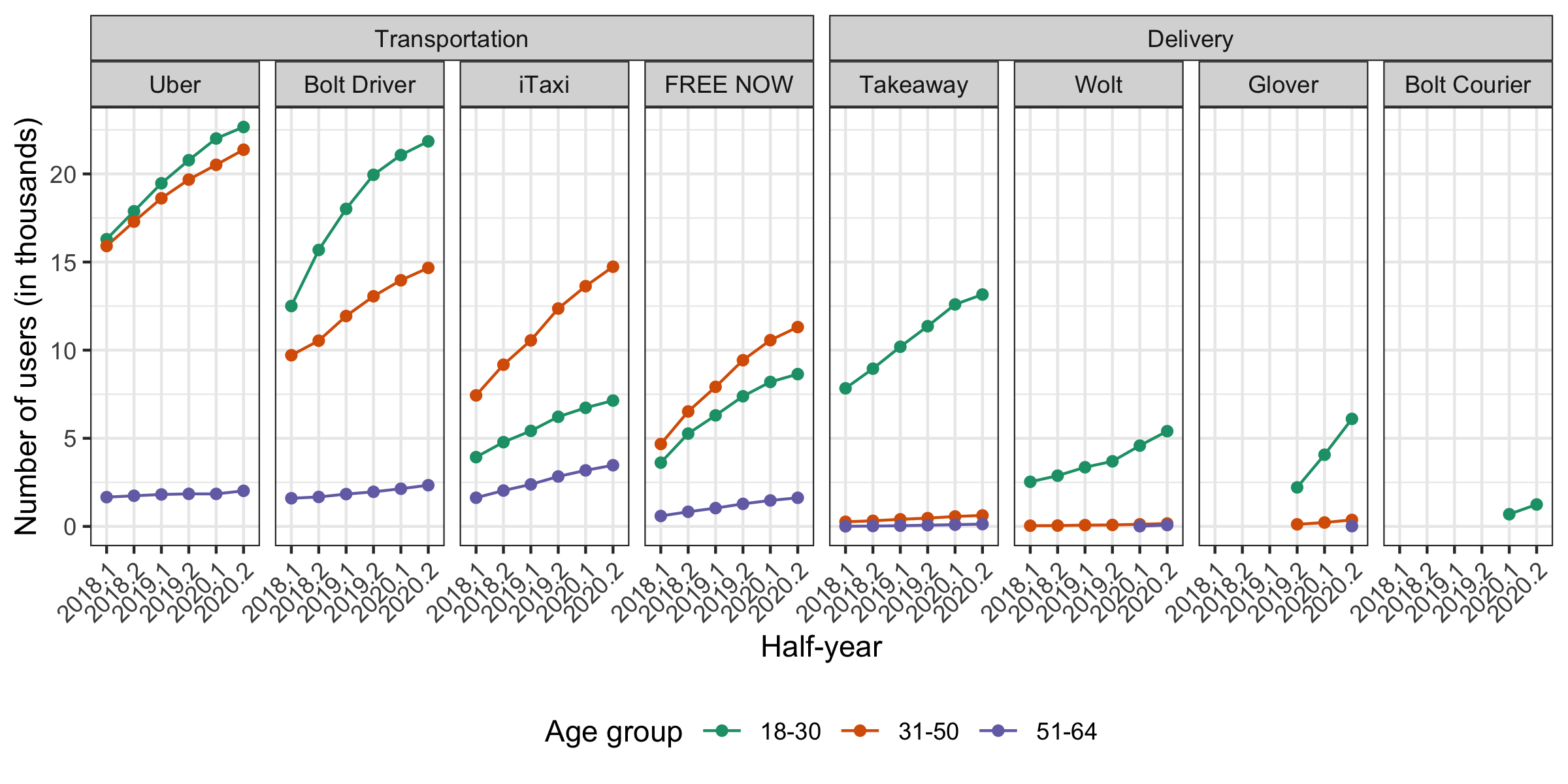}
    \includegraphics[width=0.9\textwidth]{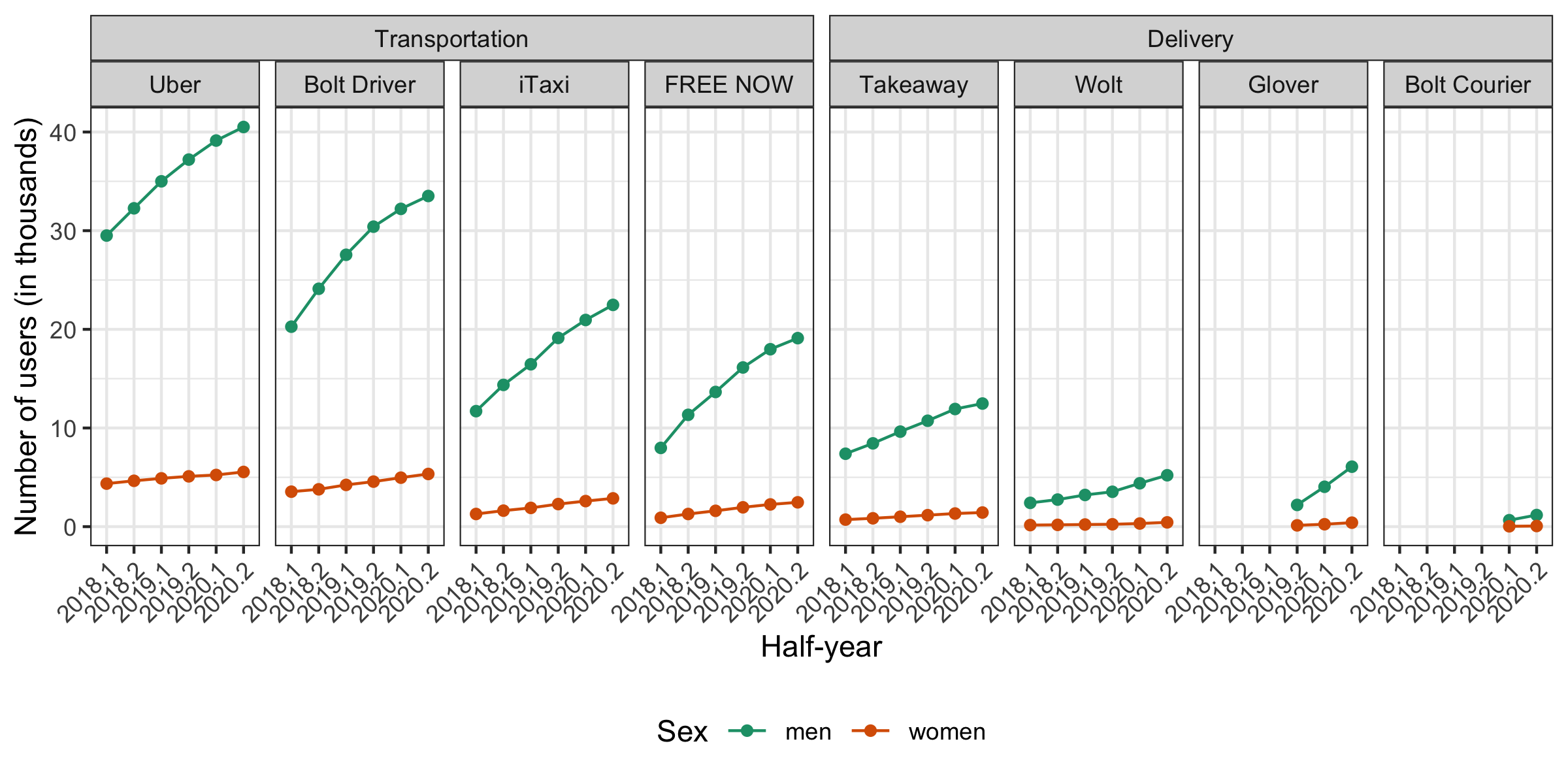}
    \caption{The monthly number of active users of selected apps in Poland by country, age and gender between 2018 and 2020}
    \label{fig-hy-country}
\end{figure}

Figure \ref{fig-hy-country} shows changes in the number of active users by nationality (top), age (middle) and sex (bottom) for different apps. The number of Polish users is significantly higher and the curve is steeper than in the case of Ukrainians and other foreigners. For each app we can observe an increase in the number of active users, which is higher than the overall increase presented in Figure \ref{fig-sel-index-total}. This suggests that existing smartphone users may have installed these apps to start providing services. We can also see a~ slight impact of COVID-19 on Glover and Wolt, but given the scope of this article we did not investigate this aspect any further.

There are also differences in the number of users of each apps when age is taken into consideration. Most users of delivery apps are aged 18--30. The highest increase can be observed for Glover and Takeaway, since these apps have the highest coverage and are the most popular. With regard to transport apps, there is a~difference in the age structure of those who use apps for licensed taxi drivers and those who use Uber and Bolt. The plurality of iTaxi users are aged 31-50 and their share grew slowly over the reference period. In contrast, for users of FREE NOW the difference between these two age groups was initially much bigger and rose faster. This may suggest that younger taxi drivers choose FREE NOW, whereas iTaxi is chosen by taxi corporations. Interesting results can be observed for Bolt and Uber. Differences between the age groups may suggest that these apps are used by different groups of users: while the trends are generally similar, the disparity between the number of users aged 18--30 and those aged 31-50 is bigger for Uber. The increase in the number of Bolt users aged 18--30 was bigger than for the other age groups but from 2019HY2 the growth slowed down.

Finally, as can be expected, most users involved in providing transportation and delivery services are male. Previous evidence for other European countries indicates that women account for about 13\% of riders, see e.g. a~study by \citet{Drahokoupil2019} for Belgium. Our results are, therefore, consistent with that proportion.  Interestingly, the share of women performing platform work is growing, especially considering the transportation sector. This trend has been recently observed and reported in Europe. Following survey data for European countries provided by \citet{Brancati2020}, platform work is becoming a~source of income for an increasing proportion of women. Importantly, differences between male and female users are closely associated with the type of tasks performed. Women are more likely to be over-represented in feminised tasks, such as translation or interactive services, while transportation and delivery services tend to be more male-dominated \citet{Brancati2020}.

Furthermore, we investigated what share of the working population (based on LFS estimates) in a~given city provided services via gig apps. Data for 19 Polish cities are compared, with the exception of the city of Sopot, where sample sizes were too small. As the LFS estimates for individual cities have relatively high variances, we decided to include confidence intervals. Details about the LFS estimates can be found in Table \ref{appen-tab-lfs-est} and information about the number of active users are presented in Table \ref{app-tab-active-cities-only} in the Appendix. 

Figure \ref{fig-hy-shares-bael} presents point estimates and 95\% confidence intervals (CI) of the ratio of the number of active users to the working population aged 16--65 for the last period, i.e. 2020HY2. It was not possible to compare transportation apps for all cities, because, while iTaxi (as well as all delivery apps) are available in all 20 cities of interest, Uber is only available in 9 cities (including Sopot), Bolt -- in 10 (including Sopot) and  FREE NOW in 8 cities (including Sopot).

\begin{figure}[ht!]
    \centering
    \includegraphics[width=\textwidth]{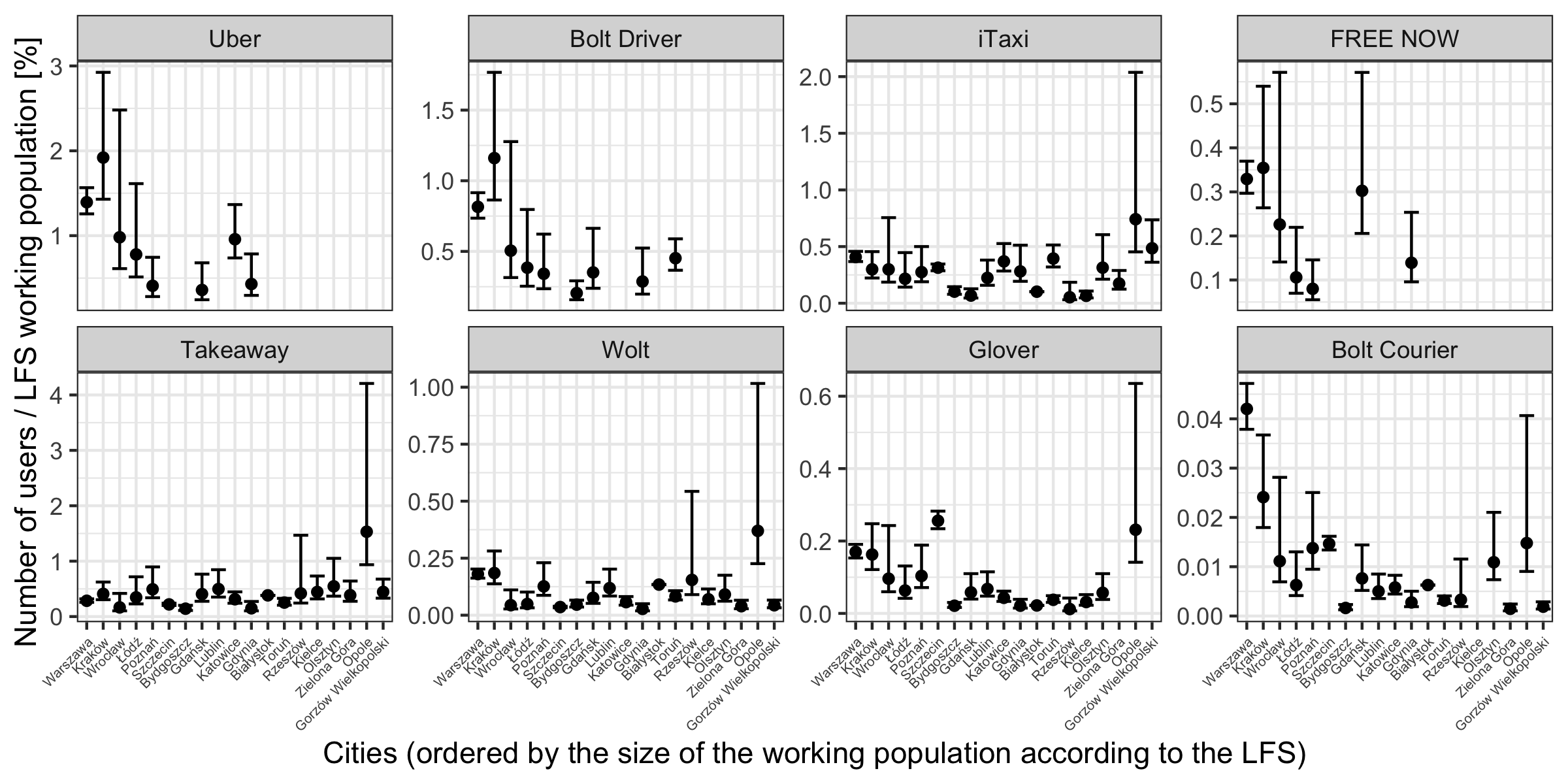}
    \caption{Point estimates and 95\% confidence intervals of the ratio of the number of active users to the LFS-based working population aged 16--65 for each app in 2020HY1. Note that the Y axis differs across the apps}
    \label{fig-hy-shares-bael}
\end{figure}

In general, the share of active users of transportation apps ranges from 0.5\% to 2\% of the working population in the selected cities. FREE NOW is the least popular one, with a~share ranging from 0.1\% to 0.5\%. While the share of app users in each city, regardless of the app, is quite similar, it is consistently the highest in Crakow and Warsaw, which could be explained by their specific character: both Crakow and Warsaw ares visited by many foreign tourists have a~large student community. The share of iTaxi users is almost the same across the cities, which may have to do with the policy of taxi corporations, which try to adjust the number of taxi drivers to a~given market, while Uber or Bolt do not impose limits on the number of users (however, they do control which drivers are matched with particular customers).

Compared to the transportation apps, the shares of delivery app users are generally smaller, even for Takeaway, the most popular app for ordering food in Poland. The highest variability can be observed for Bolt Courier, with the majority of users operating in Warsaw and Crakow. For the other apps, the pattern is similar: the share of the working age population is around 0.5\% for Takeaway and 0.1-0.2\% for Wolt and Glover. It should be noted that these figures are upper bounds, since the LFS-based working population may be underestimated as a~result of non-response among foreign-born respondents.

\begin{figure}[ht!]
    \centering
    \includegraphics[width=0.9\textwidth]{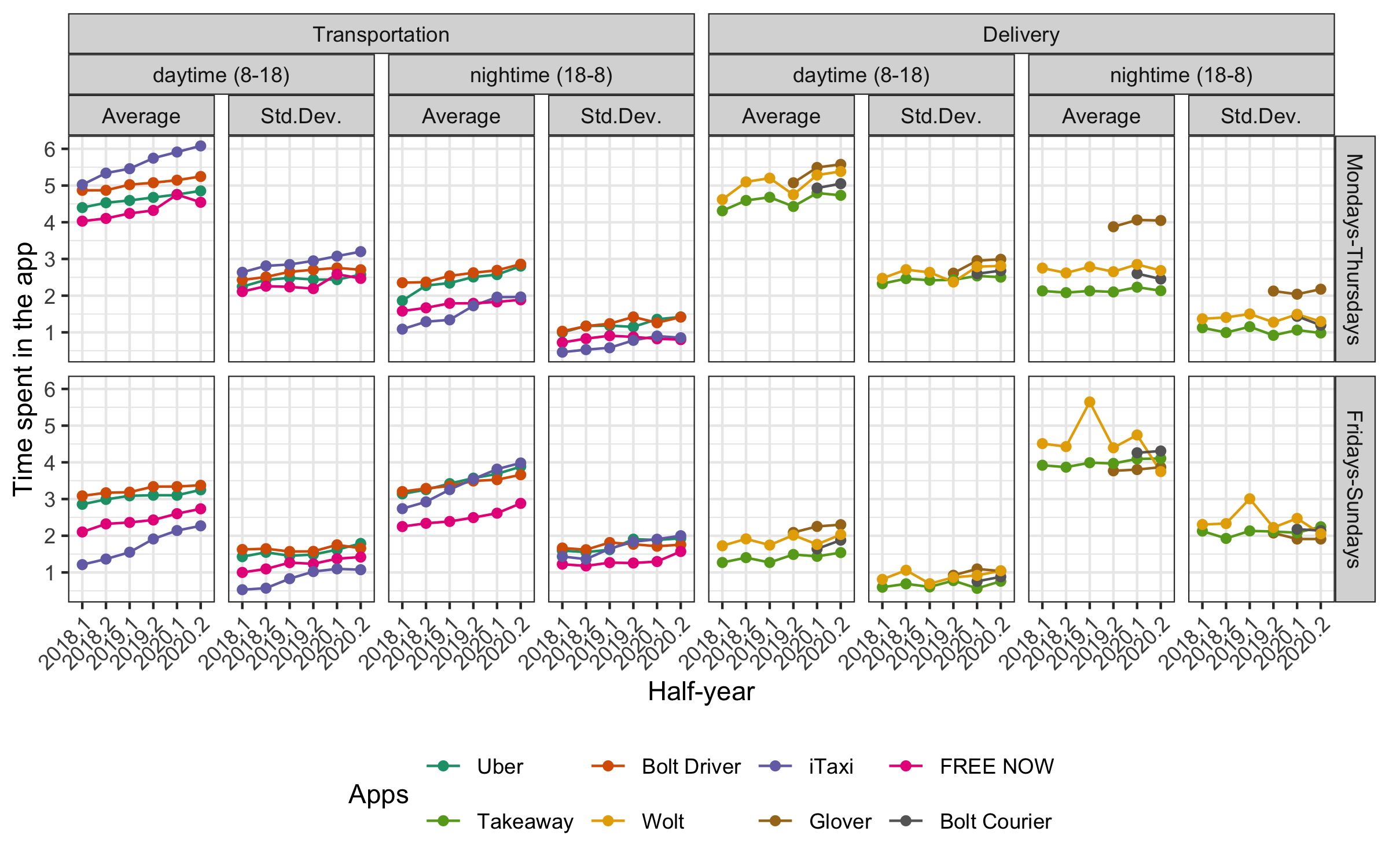}
    \caption{The mean and standard deviation (Std. Dev.) of time spent in the app by app type, app, days of the week and time of day}
    \label{fig-hy-time}
\end{figure}

Finally, we investigate the mean and standard deviation of time spent in the app during working days (Mondays-Thursdays) and at weekends (Fridays-Sundays), in the daytime (8:00--18:00) and at nighttime (18:00-8:00) over 2020 HY1. Figure \ref{fig-hy-time} presents changes in the trends for these descriptive statistics. Delivery riders tend to be more active on working days and during the day. A reverse pattern can be observed at weekends, when more intensive activity takes place at night.  This is to be expected since at weekends interest in transport services tends to increase at night, when more people are engaged in entertainment activity. However, working at night can be a~serious problem when it comes to work-life balance. As regards transportation services, the average time spent in apps on weekdays is longer during the day compared to nighttime hours. At weekends, the level of activity during the day is similar to that observed at night.

\section{Conclusions}\label{sec-conclusions}

Platform work is rapidly affecting the way labour market functions worldwide. In the broad sense, the term covers all job activities conducted by means of digital platforms and applications. Along with technology spillovers, the popularity of web-mediated activities is increasing in all economies, developing as well developed ones. As a~result, the literature concerning platform work keeps growing, providing new knowledge about this phenomenon. However, owing to data scarcity, estimates of the size of this segment on the labour market are rather limited.

In the study described in this article we made an attempt to measure the size and characteristics of the platform economy in Poland. To this end, we used big data about 22 million smartphone users collected through an advertisement system. The data enabled us to provide an upper bound of the number of drivers and couriers at very low levels of spatial aggregation. 

Our results show that the platform economy in Poland is growing. By analysing data about delivery and transportation activities performed via gig applications, we found a~growing trend between January 2018 and December 2020. As regards delivery apps, the sharpest increase was observed for Takeaway and Glover, while increases for the transport apps were more or less similar. Taking into account the demographic structure of apps users, we confirmed estimates from past studies: most platform workers are young men. Analysing the sex ratio of app users, it can be seen that the share of women is increasing  especially with respect to transport apps. By comparing the number of platform workers with the LFS estimates of the working populations in 9 largest Polish cities, we found that the share of active app users in those cities was at the level of 0.5-2\%. 



Despite the undeniable advantages of these sources, such as the fact that the data are collected passively and provide an accurate measure of time spent in the app, they are not free from certain shortcomings.  While under-coverage may not be a~serious problem, mis-classification definitely is. Not all platforms provide separate apps for workers and potential customers. This makes it difficult to identify those involved in the supply side of the gig economy. We cannot determine the threshold which should be applied to distinguish those groups. One may consider using the ILO definition of a~working person, which is based on the question about whether a~given person has worked at least one hour in the last week. But applying this rule to real-time app data may be a~challenge as there is no reference date that can be taken into account. One possible solution is to choose an arbitrary date, for example the middle of the month, but this is bound to cause errors and limit comparability with existing official statistics.

Second, mis-classification errors introduce additional bias into characteristics, such as sex, age or country of origin. All these variables are established by applying algorithms, which are not publicly available and, based on limited information disclosed by Selectivv, we are not able to determine whether their approach is correct or not. One may consider using audit samples to verify the level of errors but this would introduce sampling errors and generate additional costs.

Finally, the cost of obtaining big data is non-negligible. Companies like Selectivv collect terabytes of data over short periods of time and use cloud services to store and process them. While the price of these services is decreasing, the cost of obtaining access and the computation time required to compile target data aggregations and statistics are factors that cannot be ignored. Prior to any analyses, researchers need to specify the target population and quantities and negotiate the purchase price with these companies. This limits possibilities of data exploration as access to unit-level data is either forbidden by the company or limited by the way the data are stored.

Another problem is associated with limited access to reliable information about the reference working age population at low levels of spatial aggregation. In general, owing to small sample sizes, population surveys or coverage errors in administrative data (i.e. out-dated information, under-coverage of the foreign-born population) estimates of the size of the gig economy in these areas tend to be characterised by high MSE values. To overcome this problem, official statistics and researchers could take advantage of small area estimation, which relies on multiple sources to provide reliable estimates for areas of interest \citep[see, for example][]{van2009estimation,rao-2015}.

\printendnotes

\bibliographystyle{apalike}
\bibliography{bibliography}

\begin{thebibliography}{}

\bibitem[Abraham et~al., 2018]{Abraham2018}
Abraham, K.~G., Haltiwanger, J.~C., Sandusky, K., and Spletzer, J.~R. (2018).
\newblock {Measuring the Gig Economy: Current Knowledge and Open Issues}.

\bibitem[Aleksynska et~al., 2019]{Aleksynska2019}
Aleksynska, M., Bastrakova, A., and Kharchenko, N. (2019).
\newblock {Working Conditions on Digital Labour Platforms: Evidence from a
  Leading Labour Supply Economy}.

\bibitem[Berg, 2016]{Berg2016}
Berg, J. (2016).
\newblock {Income security in the on-demand economy: Findings and policy
  lessons from a survey of crowdworkers}.
\newblock Technical report, International Labour Office, Geneva.

\bibitem[Bohning et~al., 2017]{bohning2017capture}
Bohning, D., Van~der Heijden, P.~G., and Bunge, J. (2017).
\newblock {\em Capture-recapture methods for the social and medical sciences}.
\newblock CRC Press.

\bibitem[Bonin and Rinne, 2017]{Bonin2017}
Bonin, H. and Rinne, U. (2017).
\newblock {Omnibusbefragung zur Verbesserung der Datenlage neuer
  Besch{\"{a}}ftigungsformen}.
\newblock Technical report, IZA.

\bibitem[Brancati et~al., 2020]{Brancati2020}
Brancati, U., {C., Pesole}, A., and Fern{\'{a}}ndez-Mac{\'{i}}as, E. (2020).
\newblock {New evidence on platform workers in Europe. Results from the second
  COLLEEM survey}.
\newblock Technical report, Publications Office of the European Union,
  Luxembourg.

\bibitem[Busch, 2016]{busch2016programmatic}
Busch, O. (2016).
\newblock {\em Programmatic advertising}.
\newblock Springer, Berlin.

\bibitem[Daas et~al., 2015]{daas2015big}
Daas, P.~J., Puts, M.~J., Buelens, B., and van~den Hurk, P.~A. (2015).
\newblock Big data as a source for official statistics.
\newblock {\em Journal of Official Statistics}, 31(2):249.

\bibitem[{De Stefano} and Aloisi, 2018]{DeStefano2018}
{De Stefano}, V. and Aloisi, A. (2018).
\newblock {\em {European Legal Framework for "Digital Labour Platforms"}}.
\newblock European Commission, Luxembourg.

\bibitem[Drahokoupil and Piasna, 2019]{Drahokoupil2019}
Drahokoupil, J. and Piasna, A. (2019).
\newblock {Work in the Platform Economy: Deliveroo Riders in Belgium and the
  SMart Arrangement}.

\bibitem[Eurofound, 2018]{Eurofound2018}
Eurofound (2018).
\newblock {Employment and working conditions of selected types of platform
  work}.
\newblock Technical report, Publications Office of the European Union,
  Luxembourg.

\bibitem[Eurofound, 2019]{Eurofound2019}
Eurofound (2019).
\newblock {Platform work: Maximising the potential while safeguarding standards
  ?}
\newblock Technical report, Publications Office of the European Union,
  Luxembourg.

\bibitem[Farrell et~al., 2019]{Farrell2019}
Farrell, B.~D., Greig, F., and Hamoudi, A. (2019).
\newblock {The Evolution of the Online Platform Economy: Evidence from Five
  Years of Banking Data}.
\newblock {\em AEA Papers and Proceedings}, 109:362--366.

\bibitem[Friedman, 2014]{Friedman2014}
Friedman, G. (2014).
\newblock {Workers without employers : shadow corporations and the rise of the
  gig economy}.
\newblock {\em Review of Keynesian Economics}, 2(2):171--188.

\bibitem[Hauben et~al., 2020]{Hauben2020}
Hauben, H., Lenaerts, K., and Wayaert, W. (2020).
\newblock {The platform economy and precarious work. Publication for the
  committee on Employment and Social Affairs}.
\newblock Technical report, Policy Department for Economic, Scientific and
  Quality of Life Policies, European Parliament, Luxembourg.

\bibitem[Huws et~al., 2019]{Huws2019}
Huws, U., Spencer, N.~H., Coates, M., and Holts, K. (2019).
\newblock {The platformisation of work in Europe: results from research in 13
  European countries}.
\newblock Technical report, Foundation for European Progressive Studies, UNI
  Europa and University of Hertfordshire, Brussels.

\bibitem[ILO, 2021]{ILO2021}
ILO (2021).
\newblock {\em {World Employment and Social Outlook. The role of digital labour
  platforms in transforming the world of work}}.
\newblock International Labour Office, Geneva.

\bibitem[Koustas, 2019]{Koustas2019}
Koustas, B. D.~K. (2019).
\newblock {What Do Big Data Tell Us about Why People Take Gig Economy Jobs?}
\newblock {\em AEA Papers and Proceedings}, 109:367--371.

\bibitem[Koutsimpogiorgos et~al., 2020]{Koutsimpogiorgos2020}
Koutsimpogiorgos, N., Slageren, J.~V., Herrmann, A.~M., and Frenken, K. (2020).
\newblock {Conceptualizing the Gig Economy and Its Regulatory Problems}.
\newblock {\em Policy \& Internet}, pages 1--21.

\bibitem[Lee et~al., 2018]{Lee2018}
Lee, Z. W.~Y., Chan, T. K.~H., Balaji, M., and Chong, A. Y.-L. (2018).
\newblock {Why people participate in the sharing economy: an empirical
  investigation of Uber}.
\newblock {\em Internet Research}, 28(3):829--850.

\bibitem[Lehdonvirta, 2018]{Lehdonvirta2018}
Lehdonvirta, V. (2018).
\newblock {Flexibility in the gig economy : managing time on three online
  piecework platforms}.
\newblock {\em New Technology, Work and Employment}, 33(1):13--30.

\bibitem[Lenaerts, 2018]{Lenaerts}
Lenaerts, K. (2018).
\newblock {Industrial Relations and Social Dialogue in the Age of Collaborative
  Economy}.
\newblock Technical report, CEPS.

\bibitem[Meszmann, 2018]{Meszmann2018}
Meszmann, T.~T. (2018).
\newblock {Industrial Relations and Social Dialogue in the Age of Collaborative
  Economy. National Report Hungary}.
\newblock Technical Report~27, CELSI.

\bibitem[Mitea, 2018]{Mitea2018}
Mitea, R. D.~E. (2018).
\newblock {The Expansion of Digitally Mediated Labor: Platform-Based Economy,
  Technology-Driven Shifts in Employment, and the Novel Modes of Service Work}.
\newblock {\em Journal of Self-Governance and Management Economics},
  6(4):7--12.

\bibitem[Piasna and Drahokoupil, 2019]{Piasna2019}
Piasna, A. and Drahokoupil, J. (2019).
\newblock {Digital labour in central and eastern Europe: evidence from the ETUI
  Internet and Platform Work Survey}.

\bibitem[Rao and Molina, 2015]{rao-2015}
Rao, J. and Molina, I. (2015).
\newblock {\em Small Area Estimation}.
\newblock Wiley Series in Survey Methodology. Wiley, 2 edition.

\bibitem[Riggs et~al., 2019]{Riggs2019}
Riggs, L., Sin, I., and Hyslop, D. (2019).
\newblock {Measuring the "gig" economy: Challenges and options}.

\bibitem[Sedl{\'{a}}kov{\'{a}}, 2018]{Sedlakova2018}
Sedl{\'{a}}kov{\'{a}}, M. (2018).
\newblock {Industrial Relations and Social Dialogue in the Age of Collaborative
  Economy. National report: Slovakia}.
\newblock Technical Report~28, CELSI.

\bibitem[{Selectivv}, 2021]{selectiv2021}
{Selectivv} (2021).
\newblock {Ukrainians in Poland 2020 -- did they leave, did they come or
  stayed?}
\newblock
  \url{https://selectivv.com/en/ukrainians-in-poland-2020-did-they-leave-did-they-come-or-stayed/}.

\bibitem[Shao and Tu, 2012]{shao2012jackknife}
Shao, J. and Tu, D. (2012).
\newblock {\em The jackknife and bootstrap}.
\newblock Springer Science \& Business Media.

\bibitem[{Statistics Poland}, 2020a]{gus2020appendix}
{Statistics Poland} (2020a).
\newblock {\em {Appendix -- The foreign population in Poland during the
  COVID-19 pandemic}}.

\bibitem[{Statistics Poland}, 2020b]{gus2020ict}
{Statistics Poland} (2020b).
\newblock {\em {Information society in Poland in 2020}}.

\bibitem[{Statistics Poland}, 2020c]{gus2020bael}
{Statistics Poland} (2020c).
\newblock {\em {Labour force survey in Poland IV quarter 2020}}.

\bibitem[{Statistics Poland}, 2020d]{gus2020}
{Statistics Poland} (2020d).
\newblock {\em {The foreign population in Poland during the COVID-19
  pandemic}}.

\bibitem[Van~den Brakel and Krieg, 2009]{van2009estimation}
Van~den Brakel, J.~A. and Krieg, S. (2009).
\newblock Estimation of the monthly unemployment rate through structural time
  series modelling in a rotating panel design.
\newblock {\em Survey Methodology}, 35(2):177--190.

\end{thebibliography}

\clearpage

\appendix

\begin{center}
    \Large Supplementary materials for the paper\\
    \textit{\papertitle}
\end{center}

\section{Sample surveys -- details}

\subsection{The ICT survey}\label{appendix-ict}

The question \textit{Have you used any website or app to arrange transportation (e.g. by car) from another person in the last 12 months?} is only presented to respondents who reported using the Internet in the last year. Possible answers changed over time as shown below:

\begin{itemize}
    \item 2017 
    \begin{itemize}
        \item yes, dedicated websites or applications (e.g. UBER, BlaBlaCar)
        \item yes, other websites or applications (including social networking sites)
    \end{itemize}
    \item 2018
        \begin{itemize}
        \item yes, intermediary websites or applications dedicated specifically to organizing transport (e.g. BlaBlaCar, yanosiktls.pl, jedziemyrazem.pl)
        \item yes, other websites or applications (including social networking sites)
    \end{itemize}
    \item 2019
        \begin{itemize}
        \item yes, websites or applications that specialize in organizing trips (e.g. BlaBlaCar, jedziemyrazem.pl)
        \item yes, other websites or applications (including social networking sites)
    \end{itemize}
    \item 2020
        \begin{itemize}
        \item yes, those offered by a~given company (e.g. public transport, plane, taxi, Uber, Bolt, carsharing, electric scooters),
        \item  yes, those offered by a~private person (e.g. BlaBlaCar, jedziemyrazem.pl)
    \end{itemize}
    \item 2021
    \begin{itemize}
        \item yes, a~service offered by the company (e.g. public transport, plane, taxi, Uber, Bolt, carsharing, electric scooters),
        \item yes, a~service offered by a~private person (e.g. BlaBlaCar, jedziemyrazem.pl)
    \end{itemize}
\end{itemize}

\section{Mobile apps and websites selected for the study}\label{appendix-apps}

\begin{landscape}
	 \begin{table}[ht]
	     \centering
	     \scriptsize
	     	\caption{Gig economy apps selected for the study}
	     \label{gigeconomyapps}
	     \begin{tabular}{p{3cm}p{3.2cm}p{6cm}p{5cm}p{5cm}}
	     \hline
	       \textbf{Gig economy apps}   & \textbf{Regional availability}  & \textbf{Application process and who can apply} &  \textbf{How the user specifies working hours} & \textbf{How the user is informed about the work}\\
	     \hline       
	      Takeaway.com courier & & & & \\
	    \hline       
	    Glover & In many countries all over the world. In Poland mainly in big cities. & Creating an online account, online training is required, it is necessary to have a~vehicle (car, motorcycle, bicycle), a~smartphone with internet access, legal age is required & The courier sets their own working hours (mainly during peak demand) declaring availability (time blocks). & 15 minutes before the start of a~time block, the courier is notified to log into the app in order to confirm their availability in the booked hours. The app notifies the courier when the order has been received, .\\
	    \hline
	   Wolt courier & In 24 countries, mainly in Europe (Japan and Israel are the exception) and in 129 cities (as of 06/23/2021) & Creating an online account, it is necessary to have a~vehicle (car, motorcycle, bicycle), a~smartphone with internet access, minimum age: 16 &  The courier sets their own working hours. Working hours are flexible but mainly when restaurants are open. & After going online, the courier automatically receives orders for rides in their area. The app sends a~delivery message.\\
	    \hline
	    Bolt courier & In many countries mainly in Europe, Africa and Asia. Mexico, Ecuador and Paraguay are exceptions. & Creating an online account, it is necessary to have a~vehicle (car, motorcycle, bicycle), a~smartphone with internet access, online training is required, minimum age: 18 & The courier sets their own working hours. Working hours are flexible but mainly when restaurants are open. & After receiving a~new order, the courier receives the restaurant's address and information about how to pick up the order. The app sends a~notification. \\
	    \hline
	   
	   Uber driver & All over the world. In Poland mainly in cities.&  Valid identity card; valid driving license for at least 1 year; certificate of no criminal record; confirmation of medical and psychological examination; in Poland Uber partner drivers need to hold a~taxi license & The driver sets their working hours. & After going online, the driver automatically receives orders for rides in their area on their phone app.\\
	    \hline
	 
	   Bolt driver& Mainly in Europe in more than 150 cities. It is also available in some cities in Africa, Asia, South and North America.& An online account is required. The driver's account is activated within one working day, after all required documents have been verified. In Poland a~driver should be at least 18 years old and have at least one year of driving experience. Medical examinations and psychological tests are also required.& Drivers work according to their own schedule. No minimum hours of work are required & After going online, the driver automatically receives orders for rides in their area on their phone app.\\
	    \hline
	     \end{tabular}
	 \end{table}
\end{landscape}
	 
\begin{landscape}
	 \begin{table}[ht]
	     \centering
	     \scriptsize
	     	\caption{Non-gig economy apps selected for the study}
	     \label{nongigeconomyapps}
	     \begin{tabular}{p{3cm}p{3.2cm}p{6cm}p{5cm}p{5cm}}
	     \hline
	     	       \textbf{Non-gig economy apps}   & \textbf{Regional availability}  & \textbf{Application process and who can apply} &  \textbf{How the user specifies working hours} & \textbf{How the user is informed about the work}\\
	     \hline   
	     	   FREE NOW for drivers& FREE NOW is one of Europe's leading ride-hailing apps, which connects more than 45 million passengers with drivers in more than 150 cities (as of 06/23/2021). In Poland, the area of operation covers 7 big cities: Warsaw, Cracov, Tricity, Poznań, Wrocław, Katowice and Łódź. & Drivers wishing to join FREE NOW must register and create an account directly through the app. The next step is a~mandatory visit at the office to verify the data and present original documents (a taxi license, a driving license, car registration certificate, legalization of the taximeter, a certificate of psychological examination)& Drivers decide which days of the week and at what times they want to work. Drivers are not on duty, they can earn extra money in their spare time and combine it with working for their current corporation. & All orders are received by the driver directly via the app. The app sends notifications about new rides. Drivers can plan their day by taking orders in advance.\\
	    \hline
	    	   iTaxi Kierowca K3 & iTaxi is a~Polish technology platform, which provides travel-related services. iTaxi is available in more than 60 Polish cities. & Valid driving license; a~taxi license is required to become an iTaxi driver (minimum age: 21). Drivers have to register as self-employed. & Drivers decide which days of the week and at what times they want to work. They have to be logged into the app throughout the entire period of their availability and execute orders provided by the operator.& After going online, the driver automatically receives orders for rides in their area. The app sends notification about the new ride.\\
	    \hline
	     \end{tabular}
	 \end{table}
\end{landscape}


\clearpage
\subsection{Websites}\label{appen-websites}

For the study we identified the following websites: 

\begin{itemize}
\item http://testuj.pl
\item http://testarmy.com/
\item http://www.utest.com/
\item http://www.applause.com/
\item http://whatusersdo.com/
\item http://trymyui.com/
\item http://www.testbirds.com/
\item http://www.usertesting.com/
\item https://crowdsourcedtesting.com/
\item https://mycrowd.com/
\item https://test.io/
\item https://usabilityhub.com/
\item https://globalapptesting.com/
\item https://www.bugfinders.com/
\item https://www.fiverr.com/
\item https://www.upwork.com/
\item https://www.freelancer.pl/
\item https://www.toptal.com/
\item https://www.guru.com/
\item https://www.flexjobs.com/
\item https://www.truelancer.com/
\item https://www.clearvoice.com/ 
\item https://www.mturk.com/
\item https://www.clickworker.com/
\item https://www.microworkers.com/
\item https://www.useme.com/pl/
\end{itemize}

\clearpage

\section{Functional urban areas}\label{appen-fua}

At the time of joining the Urban Audit project, Statistics Poland did not have data on commuting. Consequently, for the purpose of 2003, 2006, 2009 and 2011 editions of the project, larger urban zones were defined as areas comprising one or more rings of LAU1 and LAU 2 units surrounding the city. 

In 2012, for cities already included in the project a~broader spatial coverage of functional urban areas was adopted (and for those introduced for the first time, FUAs were delimited by Eurostat), which was based  on results of a~survey of employment-related population flows in 2006. A functional urban area is an area in which at least 15\% of the resident population commutes to work to the city center.
 
\begin{figure}[ht!]
    \centering
    \includegraphics[width=0.8\textwidth]{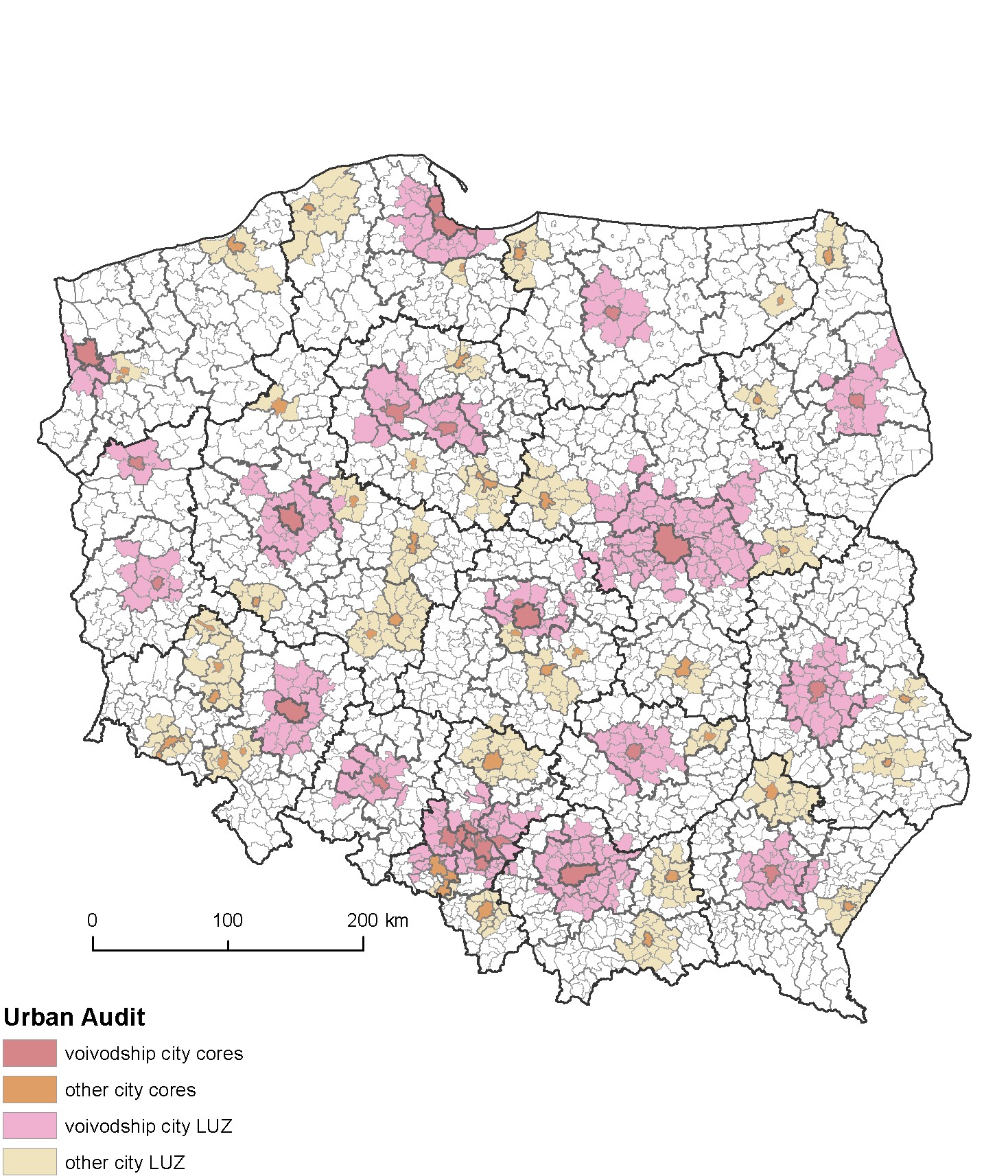}
    \caption{Functional urban areas (FUAs) in Poland in 2017. Source: \url{https://stat.gov.pl/en/regional-statistics/regional-surveys/urban-audit/larger-urban-zones-luz/}}
    \label{fig-fua-map}
\end{figure}

In the project we focused on urban cores of provincial capital cities, their functional urban areas and the remaining parts of the provinces. A list of all spatial units is presented in Table \ref{appen-tab-fua}.

\begin{table}[ht!]
    \centering
    \caption{Spatial aggregation defined in the study based on functional urban areas}
    \label{appen-tab-fua}
    \begin{tabular}{lll}
    \hline
     \textbf{Province}  & \textbf{Functional area} & \textbf{City} \\
      \hline
       Dolnośląskie  & Wrocław & Wrocław\\
       \hline
       Kujawsko-pomorskie  & Bydgoszcz & Bydgoszcz\\
        & Toruń & Toruń\\
        \hline
       Łódzkie  & Łódź & Łódź\\         
       \hline
       Lubelskie  &  Lublin &  Lublin\\         
       \hline
       Lubuskie  & Gorzów Wielkopolski &  Gorzów Wielkopolski\\     \cline{2-3}
         & Zielona Góra &  Zielona Góra\\    
         \hline
       Małopolskie  & Crakow & Crakow\\         
       \hline
       Mazowieckie   & Warsaw & Warsaw \\         
       \hline
       Opolskie   &  Opole &  Opole\\         
       \hline
       Podkarpackie   & Rzeszów & Rzeszów\\         
       \hline
       Podlaskie  &  Białystok & Białystok \\         
       \hline
       Pomorskie   &  Tricity & Gdańsk \\         
                   &   & Gdynia \\         
                   &   & Sopot \\     
                   \hline
       Śląskie  &  GZM &  Katowice\\     
                \cline{3-3}
                  &  &  13 cities of GZM\\         
                  \hline
       Świętokrzyskie  & Kielce &  Kielce\\         
       \hline
       Warmińsko-mazurskie  & Olsztyn & Olsztyn\\         
       \hline
       Wielkopolskie   &  Poznań &  Poznań\\         
       \hline
       Zachodniopomorskie    &  Szczecin & Szczecin \\         
    \hline     
    \end{tabular}
\begin{flushleft}
Note: 13 cities of the GMZ Metropolitan Area include: Bytom, Chorzów, Dąbrowa Górnicza, Gliwice, Jaworzno, Mysłowice, Piekary Śląskie, Ruda Śląska, Siemianowice Śląskie, Sosnowiec, Świętochłowice, Tychy and Zabrze.
\end{flushleft}
\end{table}

\clearpage
\section{Selected results at the level of cities, functional urban areas and provinces levels}\label{appen-city-levels}

\subsection{Details for Warsaw (capital city)}

\begin{table}[ht!]
\centering
\caption{Number of active users in Warsaw, all cities and in Poland for 2020HY2}
\label{app-tab-no-users}
\begin{tabular}{lrrrrr}
  \hline
  App &  Warsaw & All Cities & Poland & Warsaw \% of total & Cities \% of total \\ 
  \hline
  \multicolumn{6}{c}{Transportation} \\
  \hline
  Uber &  12,637 & 27,868 & 46,054 & 27.4 & 60.5 \\ 
  Bolt Driver & 7,799& 16,878 & 38,855 & 20.1 & 43.4 \\ 
  iTaxi &  3,692 & 10,024 & 25,345 & 14.6 & 39.6 \\ 
  FREE NOW &  2,954 & 5,692 & 21,573 & 13.7 & 26.4 \\
  \hline
  \multicolumn{6}{c}{Delivery} \\
  \hline
  Takeaway &  1,878 & 13,563 & 13,910 & 13.5 & 97.5 \\ 
  Wolt & 1,956& 5,060 & 5,645 & 34.7 & 89.6 \\ 
  Glover &  2,358 & 5,906 & 6,482 & 36.4 & 91.1 \\ 
  Bolt Courier & 533 & 1,168 & 1,243 & 42.9 & 94.0 \\ 
   \hline
\end{tabular}
\begin{flushleft}
Note: Warsaw \% of total is calculated as Warsaw to Total, Cities \% of total is calculated as All Cities to Total. All Cities include those listed in Table \ref{appen-tab-fua} except for the 13 cities of GMZ.
\end{flushleft}
\end{table}

\clearpage

\subsection{Detailed information about all cities in 2020HY2}

\begin{table}[ht]
\centering
\caption{Number of active users in cities for 2020HY2}
\label{app-tab-active-cities-only}
\begin{tabular}{lrrrrrrrr}
  \hline
  & \multicolumn{4}{c}{Transport} & \multicolumn{4}{c}{Delivery} \\ 
  \cline{2-9}
 City & Uber & Bolt & iTaxi & FREENOW & Takeaway & Wolt & Glover & Bolt  \\ 
  \hline
  Białystok & -- & -- & 96 & -- & 630 & 205 & 56 & 24 \\ 
  Bydgoszcz & -- & 274 & 178 & -- & 297 & 101 & 71 & 12 \\ 
  Gdańsk & 472 & 542 & 104 & 450 & 778 & 156 & 170 & 32 \\ 
  Gdynia & 479 & 401 & 373 & 120 & 245 & 48 & 53 & 10 \\ 
  Gorzów Wielkopolski & -- & -- & 178 & -- & 311 & 27 & 15 & 5 \\ 
  Katowice & 1578 & -- & 482 & -- & 558 & 106 & 121 & 21 \\ 
  Kielce & -- & -- & 12 & -- & 442 & 75 & 65 & 18 \\ 
  Crakow & 6808 & 4166 & 994 & 1219 & 1619 & 792 & 893 & 170 \\ 
  Łódź & 1762 & 793 & 405 & 230 & 908 & 140 & 261 & 33 \\ 
  Lublin & -- & -- & 344 & -- & 958 & 226 & 187 & 18 \\ 
  Olsztyn & -- & -- & 190 & -- & 533 & 90 & 122 & 20 \\ 
  Opole & -- & -- & 310 & -- & 912 & 238 & 204 & 30 \\ 
  Poznań & 835 & 701 & 506 & 64 & 1288 & 350 & 410 & 62 \\ 
  Rzeszów & -- & -- & 41 & -- & 459 & 169 & 36 & 22 \\ 
  Sopot & 80 & 249 & 179 & 35 & 196 & 38 & 42 & 8 \\ 
  Szczecin & -- & -- & 461 & -- & 387 & 52 & 254 & 55 \\ 
  Toruń & -- & 336 & 370 & -- & 276 & 94 & 66 & 11 \\ 
  Warsaw & 12637 & 7799 & 3692 & 2954 & 1878 & 1956 & 2358 & 533 \\ 
  Wrocław & 3217 & 1617 & 931 & 620 & 540 & 165 & 505 & 78 \\ 
  Zielona Góra & -- & -- & 178 & -- & 348 & 32 & 17 & 6 \\ 
   \hline
\end{tabular}
\end{table}

\clearpage
\subsection{Plots for selected cities, their FUAs and provinces}\label{appen-cities-fuas}

\begin{figure}[ht!]
    \centering
    \includegraphics[width=\textwidth]{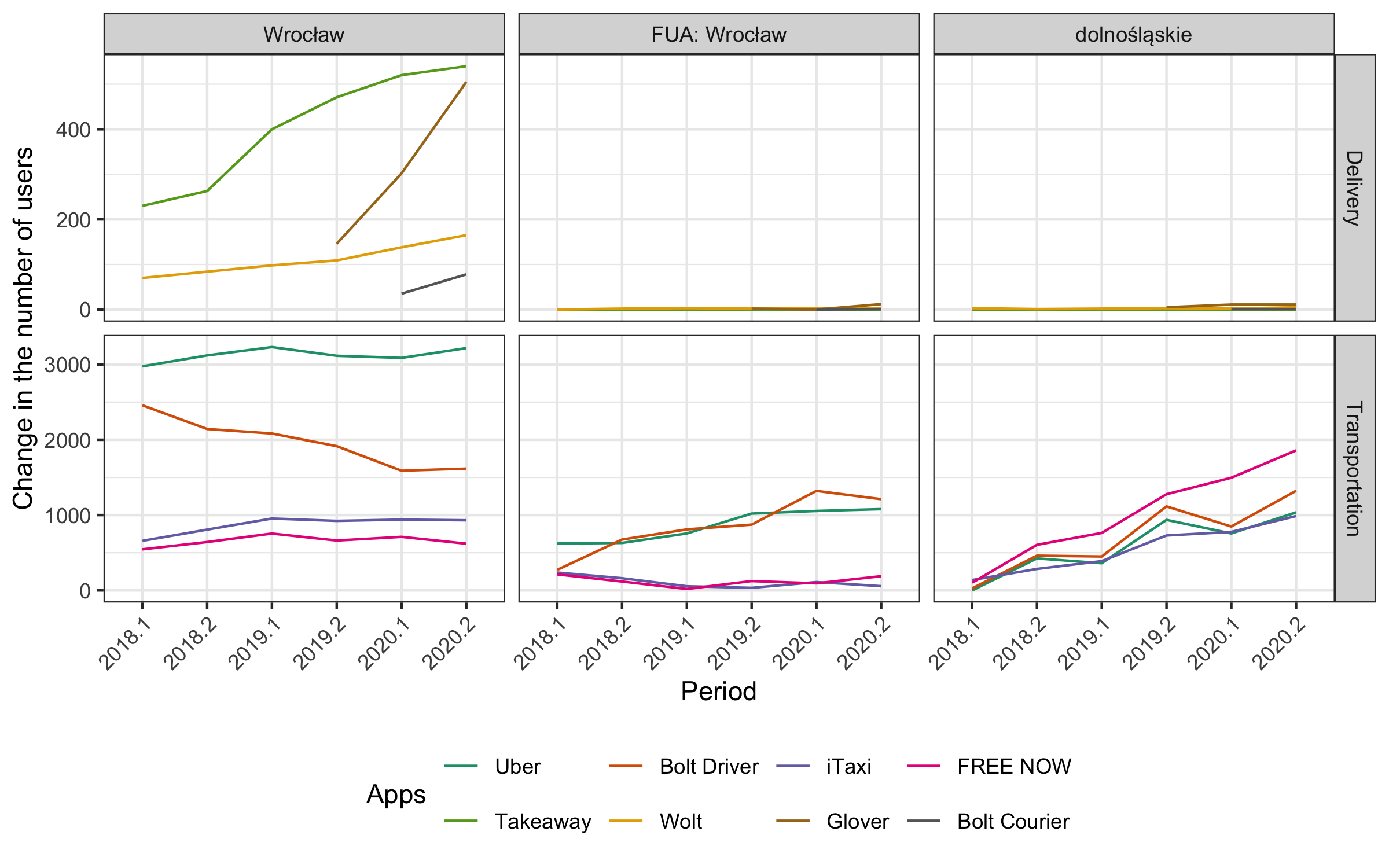}
    \caption{Changes in the number of active users for selected apps in Wrocław, its FUA and the whole province by app category between 2018 and 2020}
    \label{fig-annex-1}
\end{figure}

\begin{figure}[ht!]
    \centering
    \includegraphics[width=\textwidth]{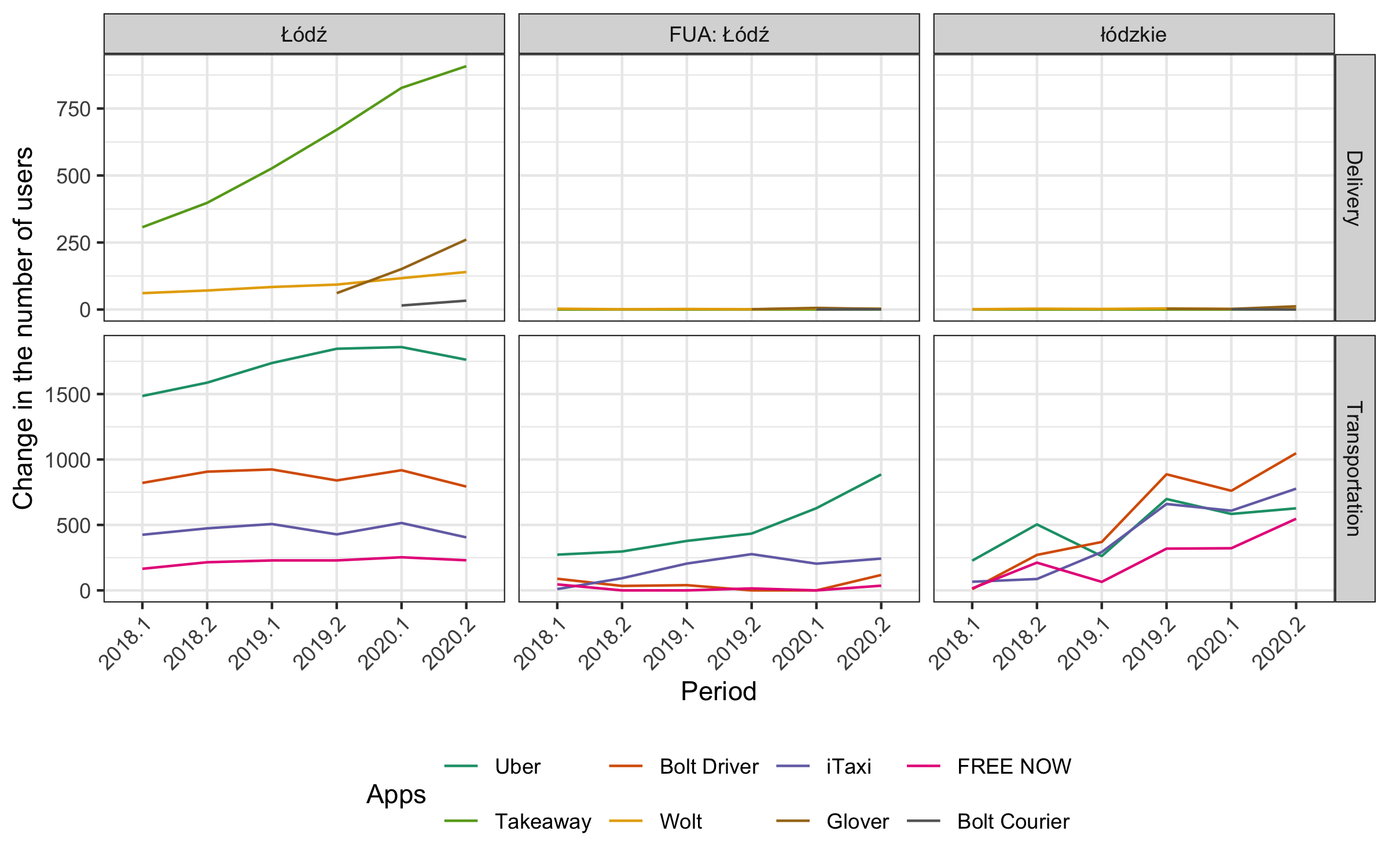}
    \caption{Change in the number of active users for selected apps in Łódź, its FUA and the whole province by type between 2018 and 2020}
    \label{fig-annex-3}
\end{figure}

\begin{figure}[ht!]
    \centering
    \includegraphics[width=\textwidth]{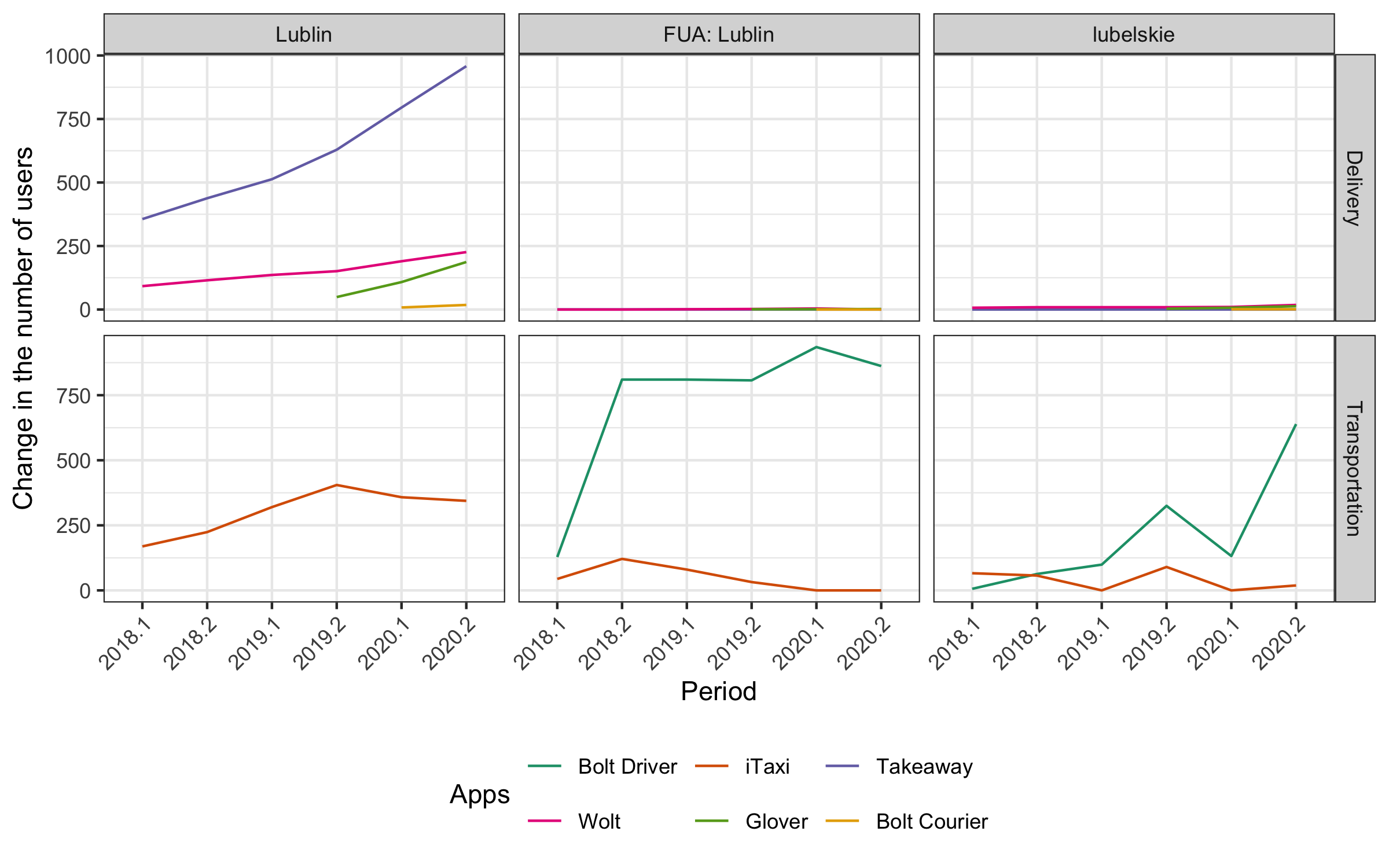}
    \caption{Change in the number of active users for selected apps in Lublin, its FUA and the whole province by type between 2018 and 2020}
    \label{fig-annex-4}
\end{figure}

\begin{figure}[ht!]
    \centering
    \includegraphics[width=\textwidth]{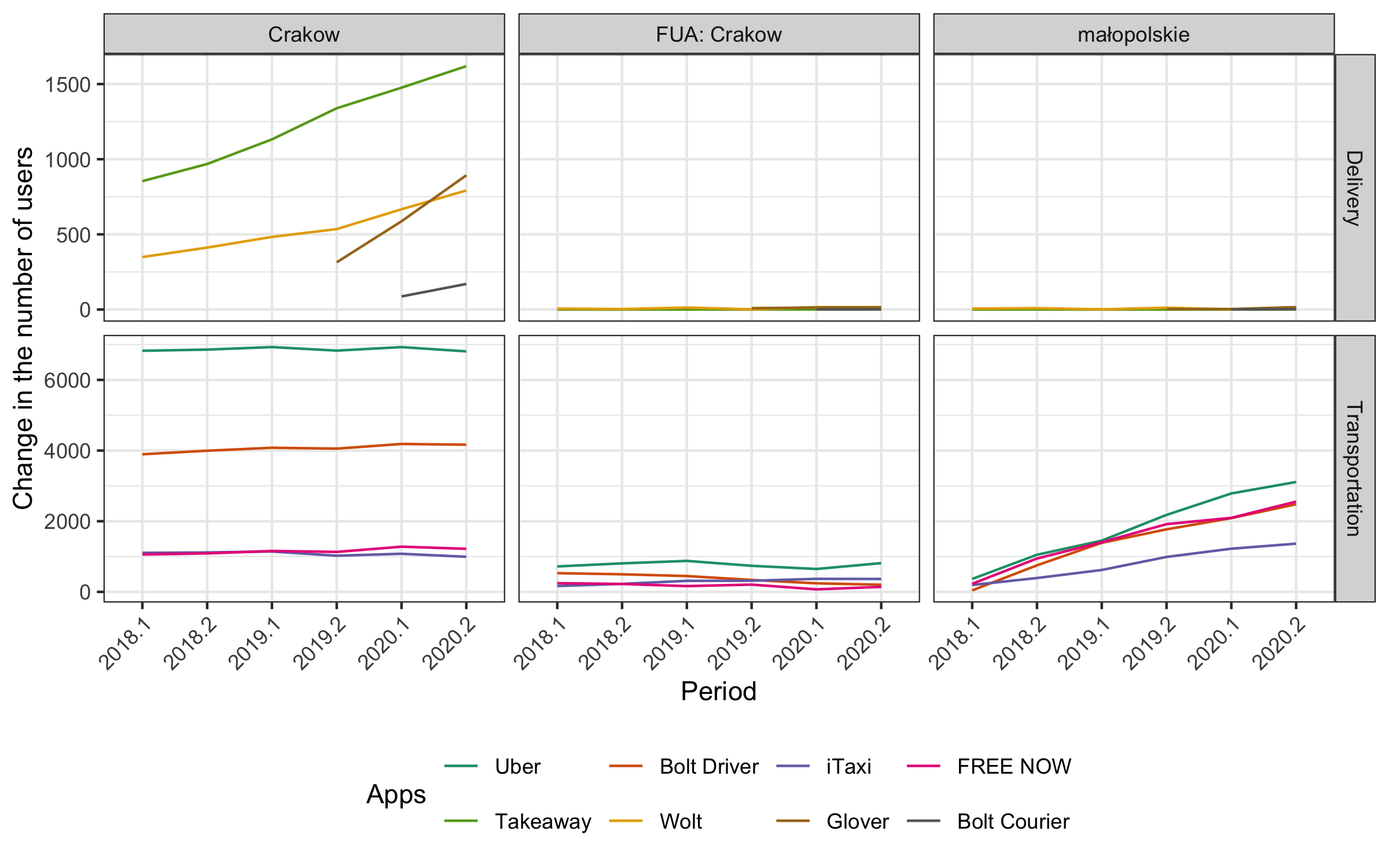}
    \caption{Change in the number of active users for selected apps in Crakow, its FUA and the whole province by type between 2018 and 2020}
    \label{fig-annex-6}
\end{figure}

\begin{figure}[ht!]
    \centering
    \includegraphics[width=\textwidth]{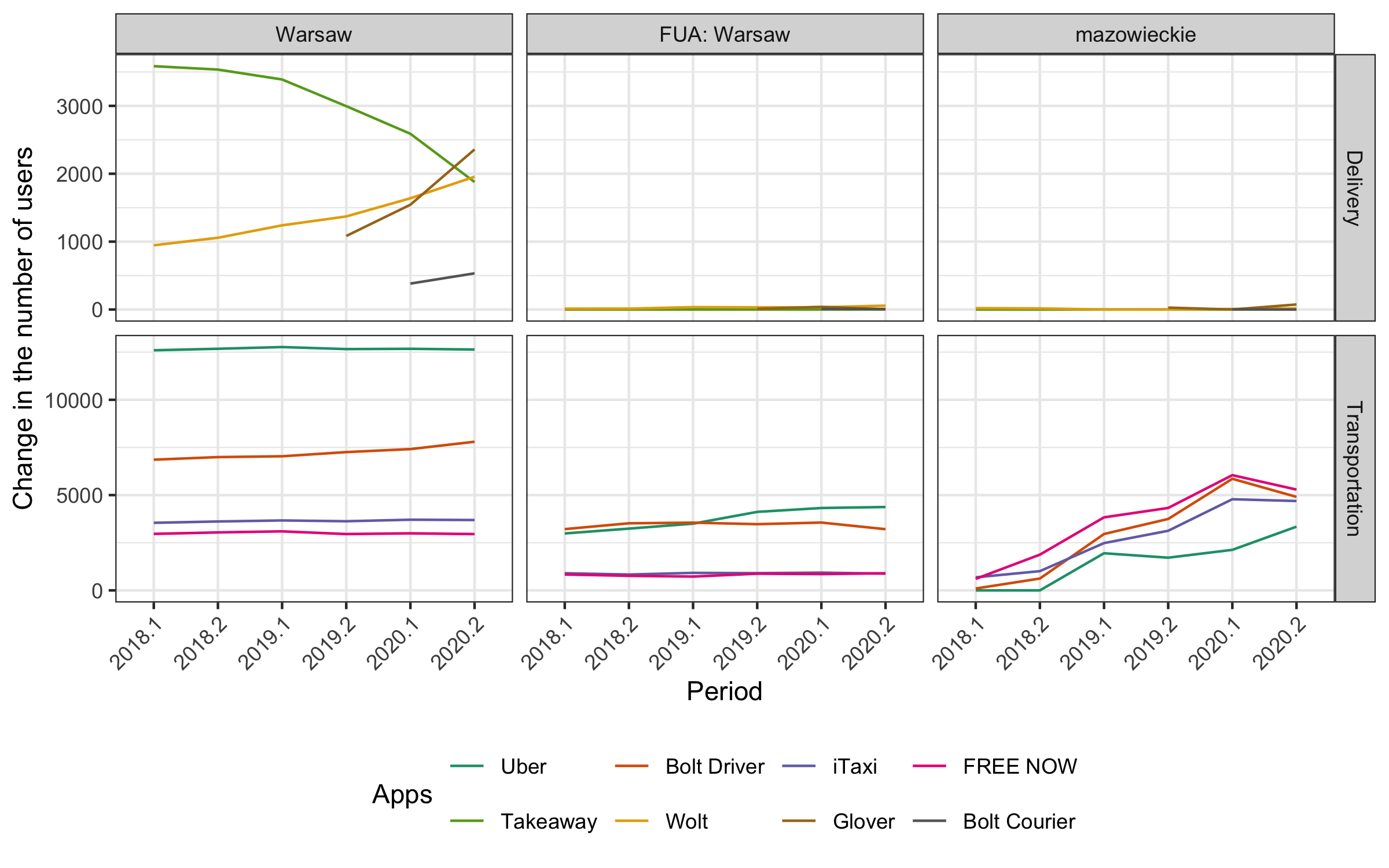}
    \caption{Change in the number of active users for selected apps in Warsaw, its FUA and the whole province by type between 2018 and 2020}
    \label{fig-annex-7}
\end{figure}

\begin{figure}[ht!]
    \centering
    \includegraphics[width=\textwidth]{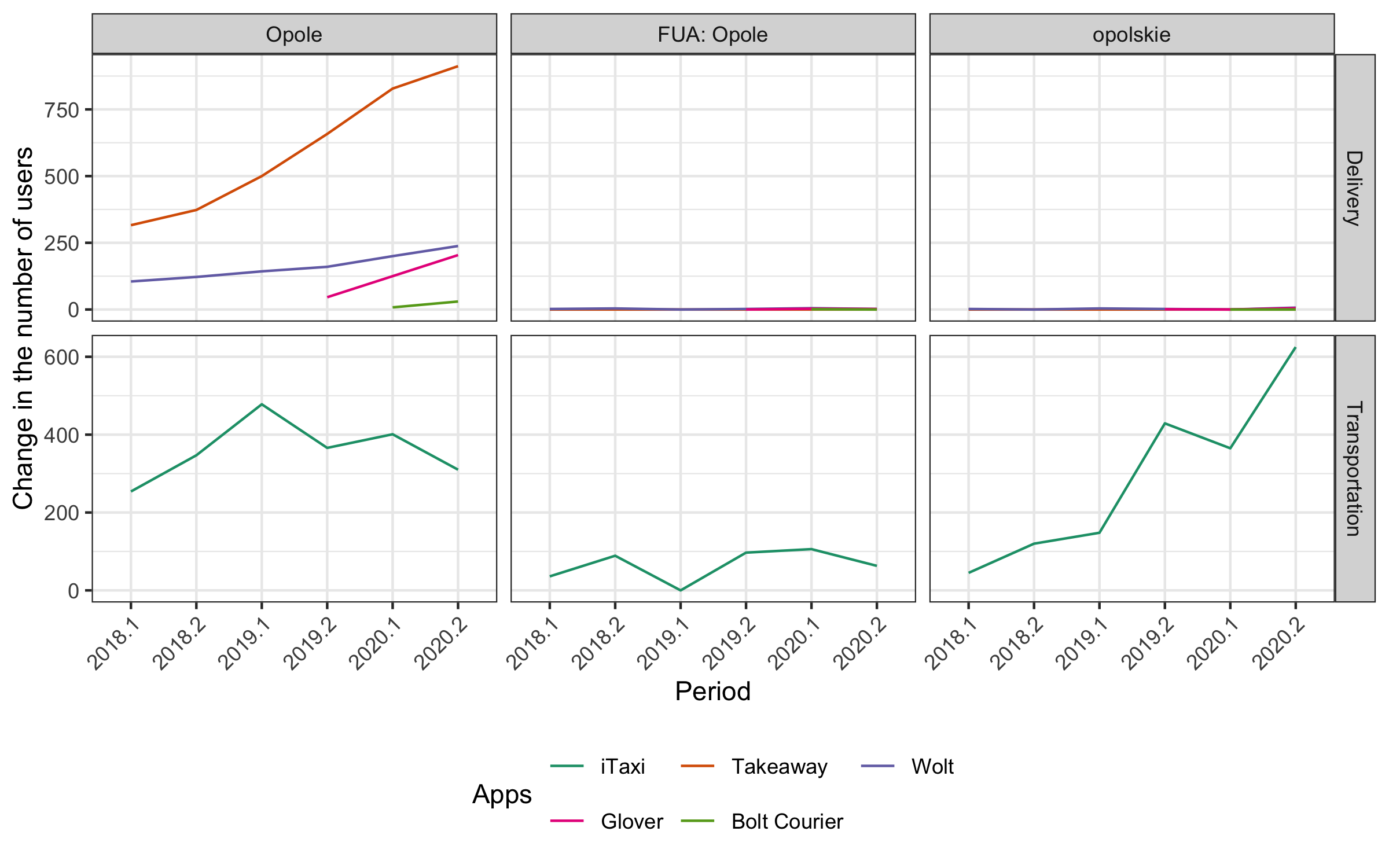}
    \caption{Change in the number of active users for selected apps in Opole, its FUA and voivodeship by type and between 2018 and 2020}
    \label{fig-annex-8}
\end{figure}

\begin{figure}[ht!]
    \centering
    \includegraphics[width=\textwidth]{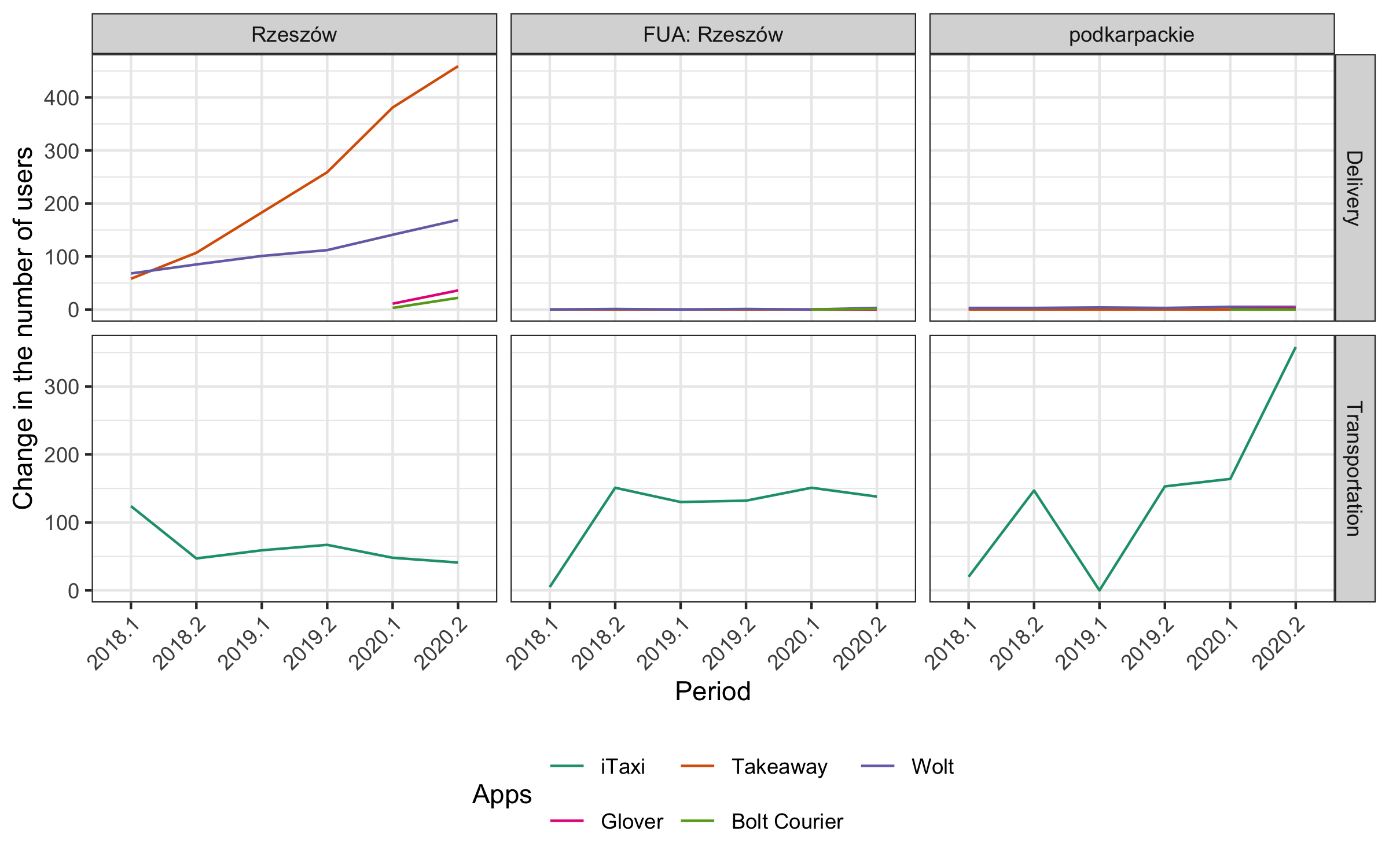}
    \caption{Change in the number of active users for selected apps in Rzeszów, its FUA and the whole province by type between 2018 and 2020}
    \label{fig-annex-9}
\end{figure}

\begin{figure}[ht!]
    \centering
    \includegraphics[width=\textwidth]{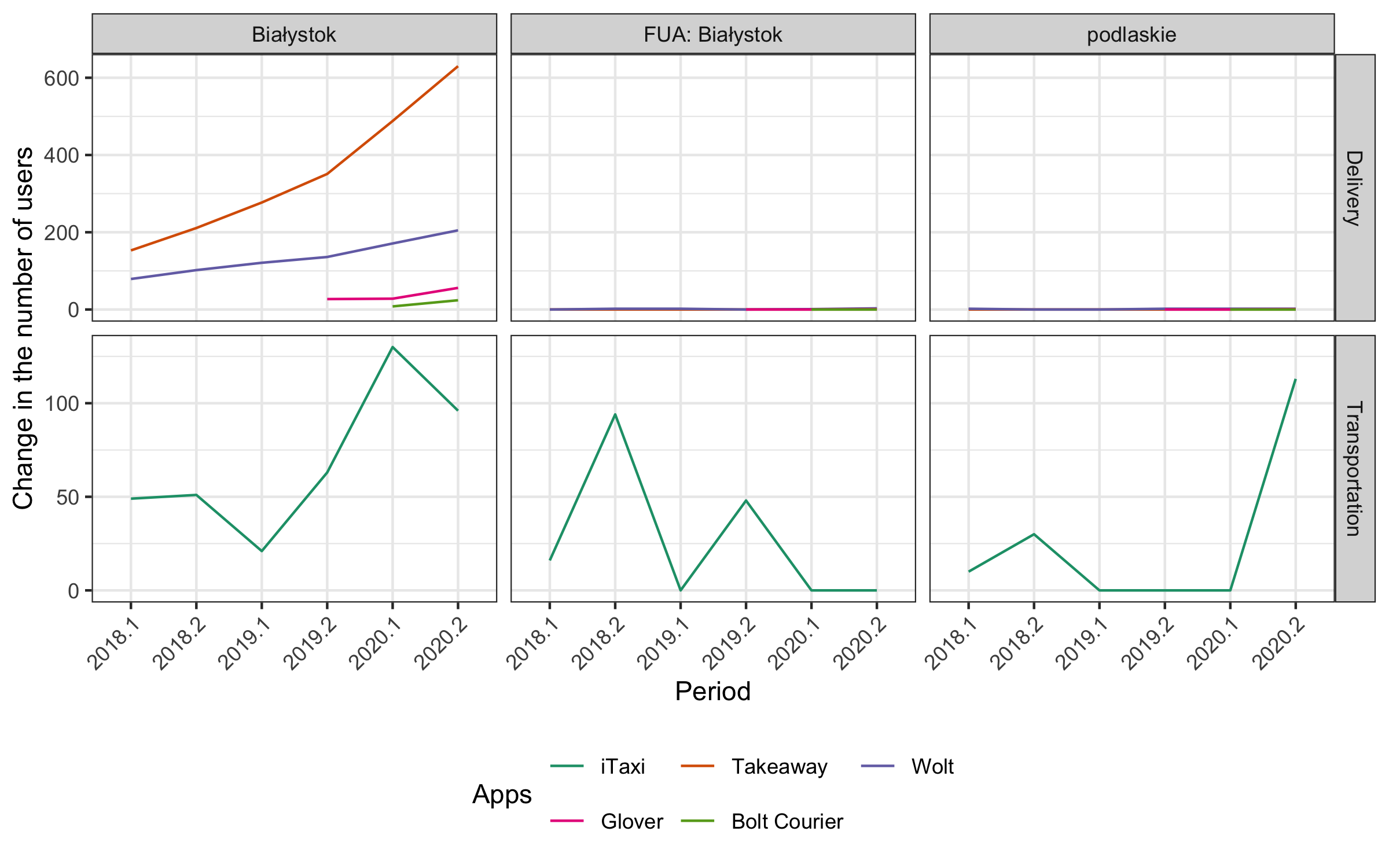}
    \caption{Change in the number of active users for selected apps in Białystok, its FUA and the whole province by type between 2018 and 2020}
    \label{fig-annex-10}
\end{figure}

\begin{figure}[ht!]
    \centering
    \includegraphics[width=\textwidth]{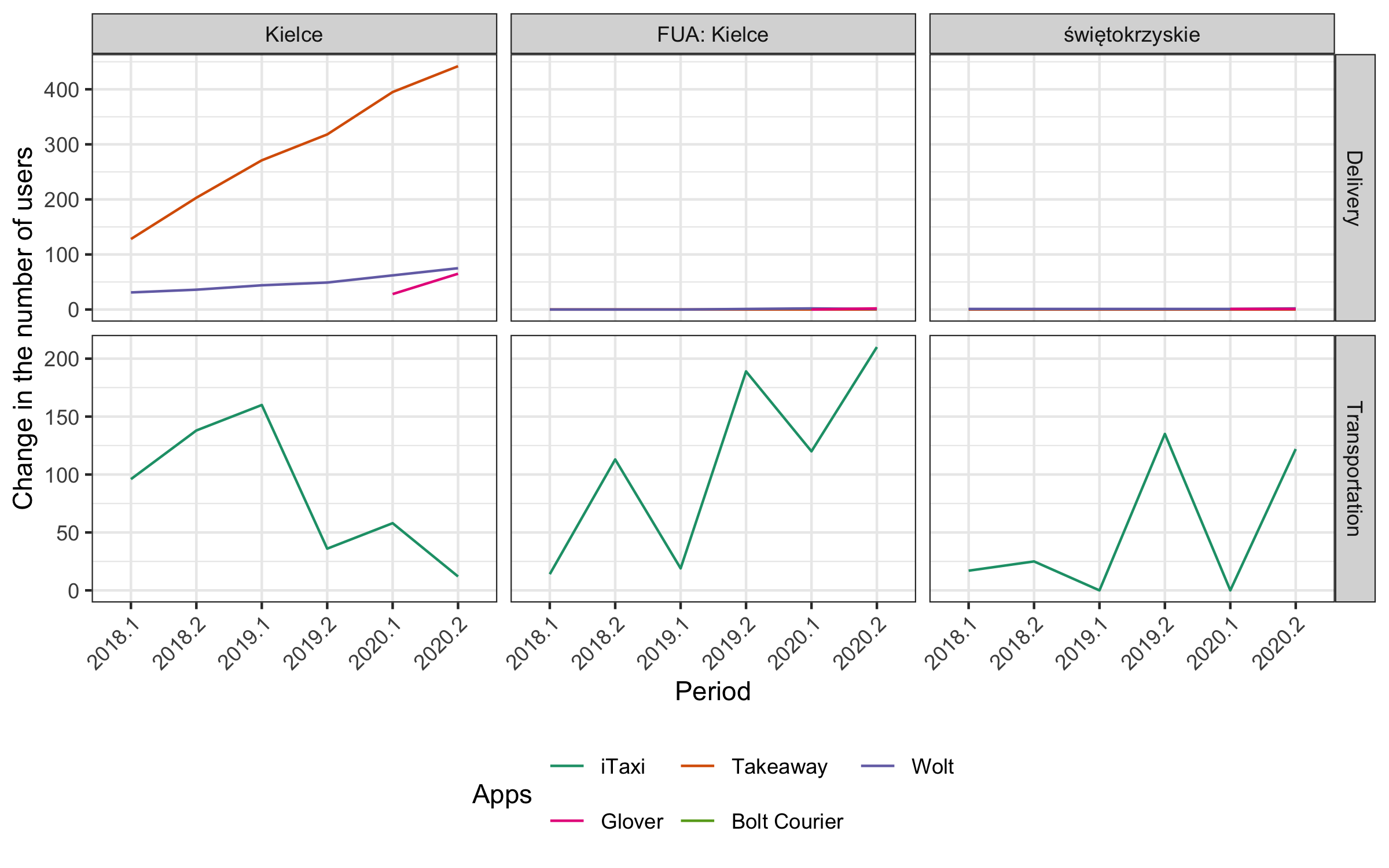}
    \caption{Change in the number of active users for selected apps in Kielce, its FUA and the whole province by type between 2018 and 2020}
    \label{fig-annex-13}
\end{figure}

\begin{figure}[ht!]
    \centering
    \includegraphics[width=\textwidth]{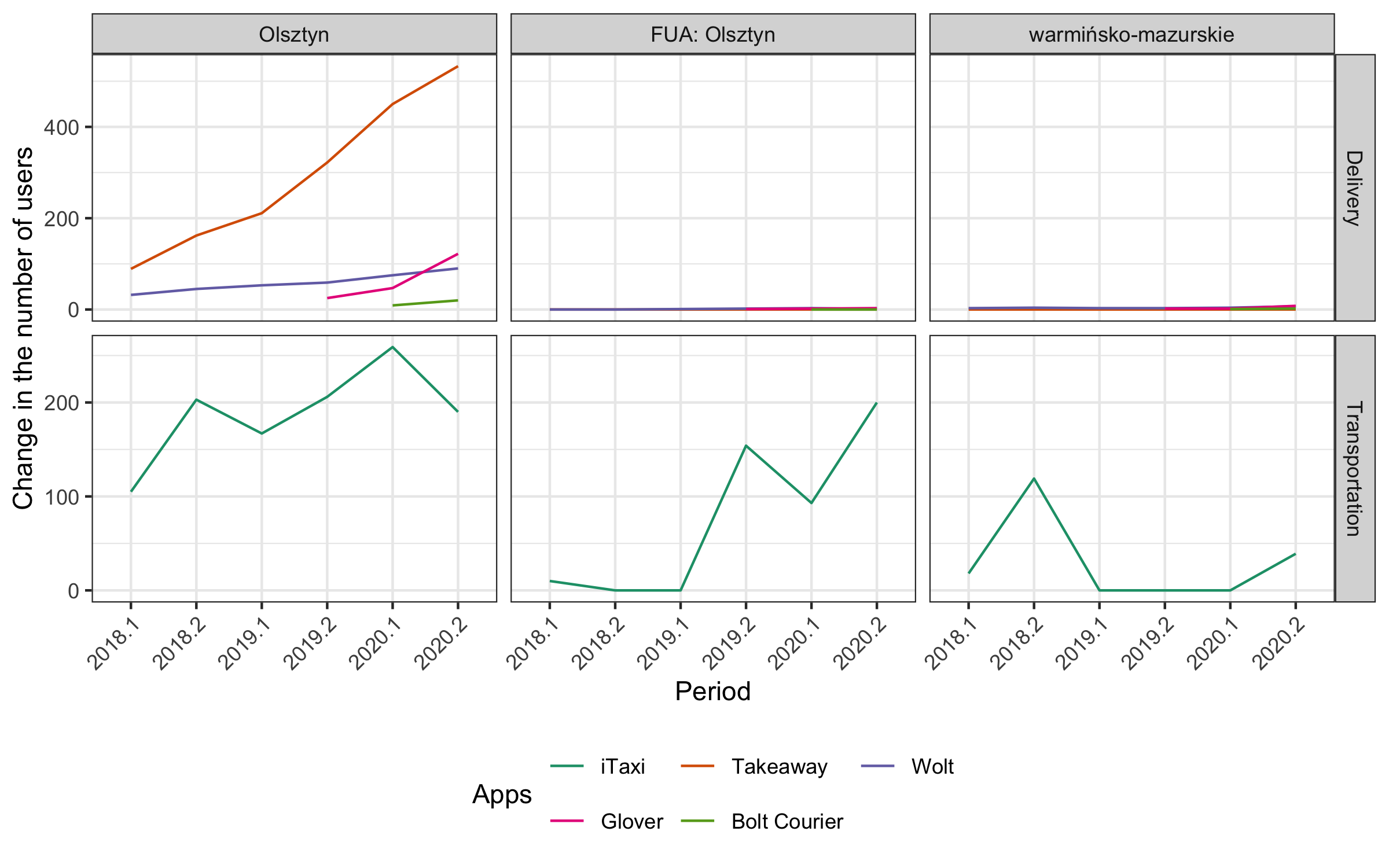}
    \caption{Change in the number of active users for selected apps in Olsztyn, its FUA and the whole province by type between 2018 and 2020}
    \label{fig-annex-14}
\end{figure}

\begin{figure}[ht!]
    \centering
    \includegraphics[width=\textwidth]{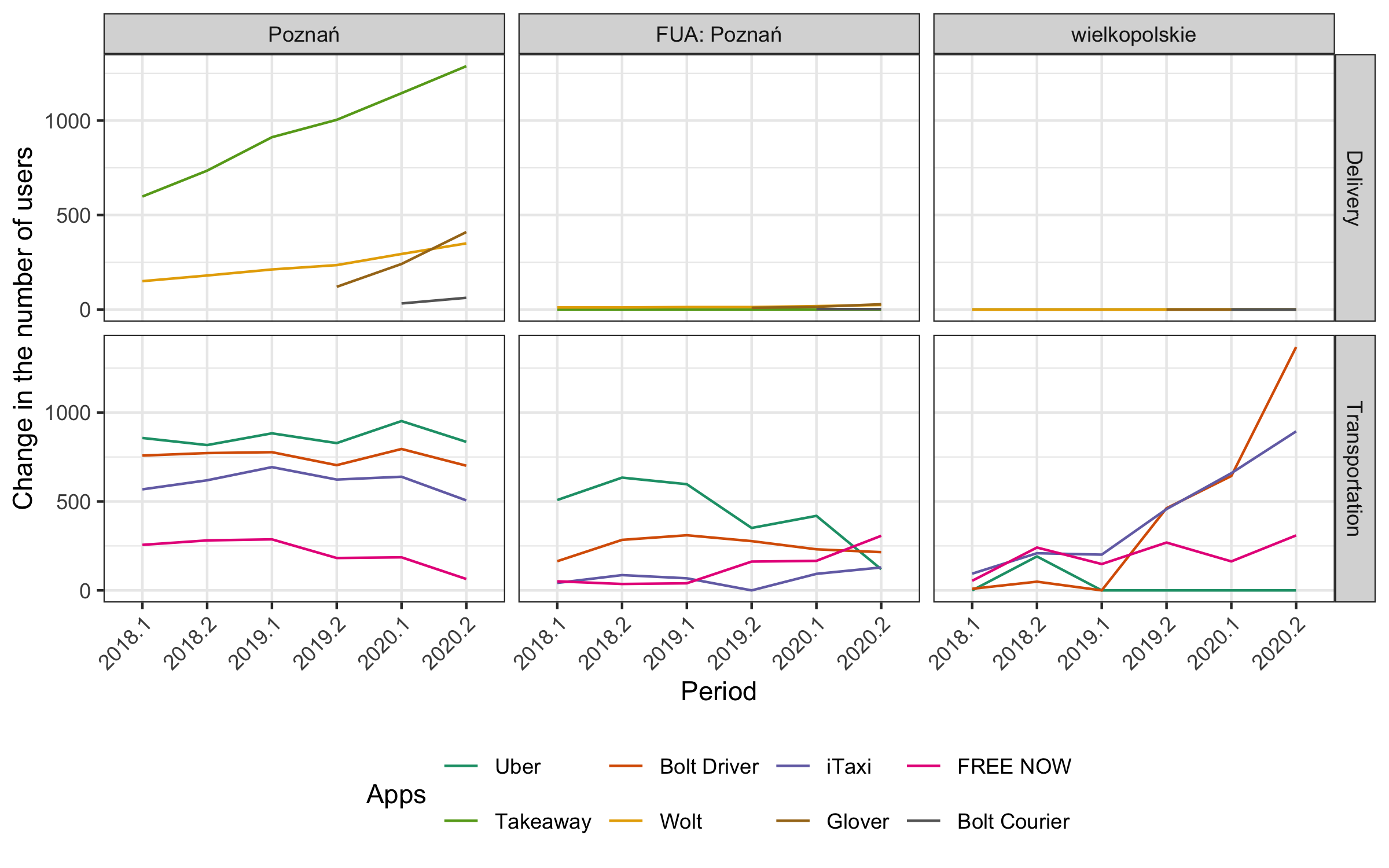}
    \caption{Change in the number of active users for selected apps in Poznań, its FUA and the whole province by type between 2018 and 2020}
    \label{fig-annex-15}
\end{figure}

\begin{figure}[ht!]
    \centering
    \includegraphics[width=\textwidth]{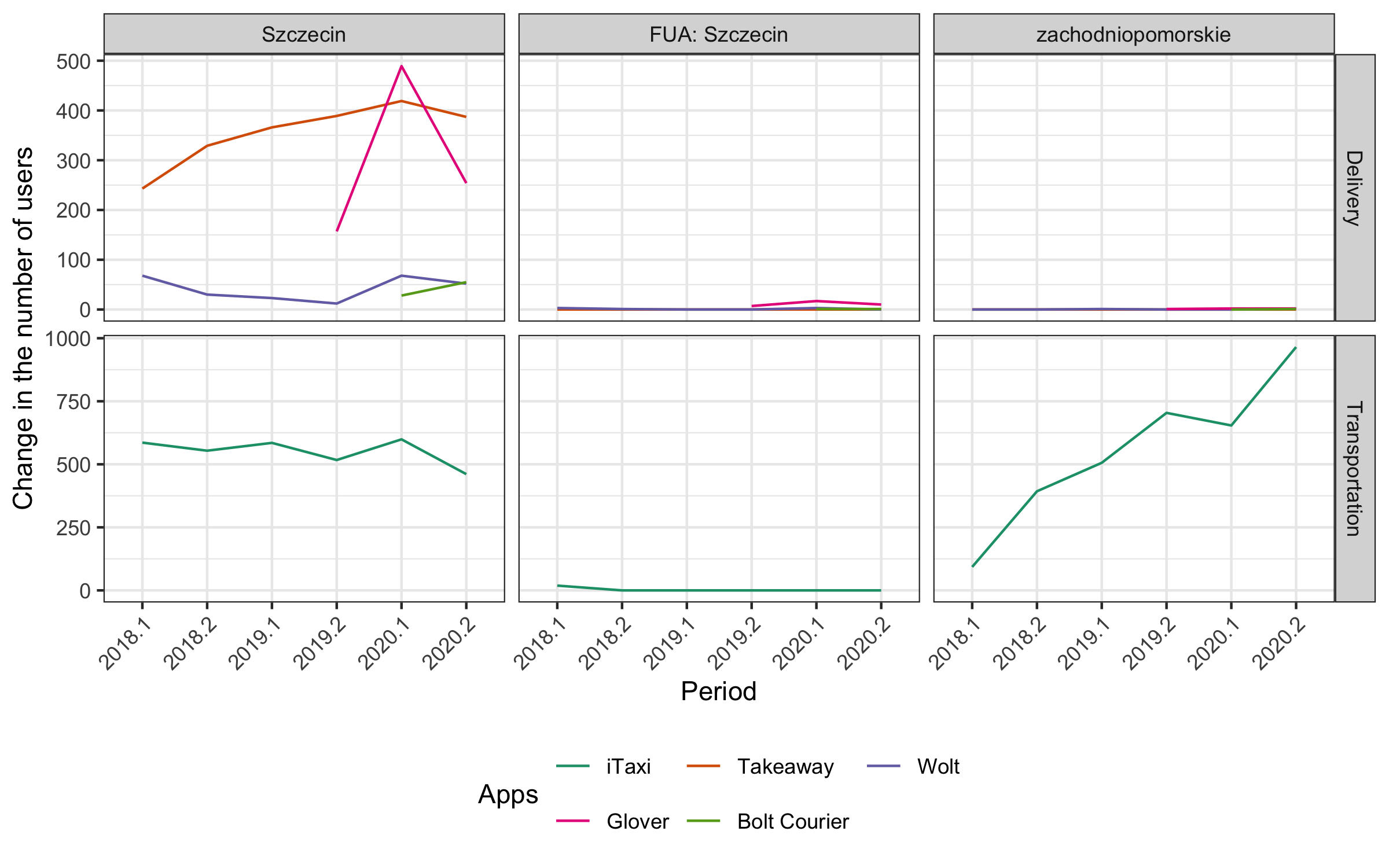}
    \caption{Change in the number of active users for selected apps in Szczecin, its FUA and the whole province by type between 2018 and 2020}
    \label{fig-annex-16}
\end{figure}

\clearpage

\section{Labour Force Survey}\label{appendix-lfs} 

LFS results are generalised for the population using calibration weights, which are constructed in three steps. The purpose of the first step is to compensate for disproportions in the sample structure. This is done by calculating primary weights of dwellings (treated as households), which are defined as reciprocals of their selection probabilities. 

During the second step, refusals are taken into account by calculating response rates for 6 different size categories of localities in each province. The response rates are defined as ratios of the sum of primary weights of surveyed households to the sum of primary weights of households that qualified to be surveyed.  Final household weights are obtained by multiplying primary weights by reciprocals of corresponding response rates. 

The goal of the third step is to ensure consistency with demographic estimates. This is achieved by adjusting secondary weights of surveyed respondents (final household weights assigned to individual household members) so that they correctly sum up to population counts in 48 categories cross-classified by the place of residence (urban/rural), sex and 12 age groups in each province. To this end, within each cross-classification a factor is calculated, which is given as the ratio of the population count in each cross-classification (known from current demographic estimates) to the sum of secondary weights of respondents in a given cross-classification.  Final weights of respondents are obtained by multiplying secondary weights of respondents by an appropriate factor.

The variance of estimators in the LFS is estimated by using the bootstrap method. A detailed description of the approach applied to complex two-stage sampling designs, can be found in the monograph by \citet{shao2012jackknife}.

\begin{landscape}
\begin{table}[ht]
\caption{Direct estimates,their standard errors (in thousands) and sample sizes of the working population size (aged 18--64) based on the LFS data from between 2018HY1 to 2020HY}
\label{appen-tab-lfs-est}
\centering
\begin{tabular}{l|rrr|rrr|rrr|rrr|rrr}
  \hline
  & \multicolumn{3}{c}{2018HY1} & \multicolumn{3}{c}{2018HY2} & \multicolumn{3}{c}{2019HY1} & 
  \multicolumn{3}{c}{2019HY2} & \multicolumn{3}{c}{2020HY1} \\ 
 City & $\hat{N}$ & $SE(\hat{N})$ & n & 
        $\hat{N}$ & $SE(\hat{N})$ & n & 
        $\hat{N}$ & $SE(\hat{N})$ & n & 
        $\hat{N}$ & $SE(\hat{N})$ & n & 
        $\hat{N}$ & $SE(\hat{N})$ & n \\ 
  \hline
Białystok & 132.2 & 2.9 & 988 & 132.4 & 11.5 & 934 & 128.2 & 0.6 & 892 & 138.9 & 1.6 & 879 & 127.4 & 0.1 & 845 \\ 
  Bydgoszcz & 175.9 & 20.2 & 1359 & 170.4 & 24.0 & 1250 & 165.5 & 13.4 & 1179 & 163.3 & 20.2 & 1046 & 183.6 & 27.9 & 1052 \\ 
  Gdańsk & 180.5 & 33.0 & 847 & 190.5 & 42.3 & 842 & 207.0 & 16.4 & 852 & 206.6 & 36.4 & 807 & 170.3 & 40.9 & 660 \\ 
  Gdynia & 142.7 & 3.9 & 363 & 145.1 & 14.4 & 351 & 134.8 & 8.1 & 326 & 132.4 & 6.9 & 292 & 145.3 & 33.5 & 319 \\ 
  Gorzów Wielkopolski & 67.0 & 9.8 & 1002 & 59.0 & 13.8 & 809 & 63.4 & 5.9 & 848 & 53.9 & 11.7 & 722 & 52.8 & 9.2 & 713 \\ 
  Katowice & 139.1 & 29.5 & 819 & 142.3 & 20.4 & 725 & 150.3 & 16.9 & 705 & 167.4 & 37.0 & 690 & 155.5 & 23.7 & 661 \\ 
  Kielce & 84.7 & 16.5 & 869 & 88.9 & 22.6 & 870 & 86.4 & 19.0 & 884 & 86.8 & 29.6 & 867 & 89.0 & 17.9 & 807 \\ 
  Crakow & 408.0 & 15.3 & 904 & 418.3 & 20.3 & 793 & 415.6 & 63.3 & 776 & 430.5 & 53.5 & 756 & 360.8 & 63.2 & 640 \\ 
  Lublin & 176.0 & 20.7 & 1371 & 167.6 & 19.2 & 1195 & 162.8 & 26.5 & 1156 & 153.0 & 30.4 & 1033 & 160.0 & 33.7 & 1022 \\ 
  Łódź & 328.5 & 16.1 & 637 & 334.3 & 16.9 & 510 & 291.5 & 1.8 & 489 & 298.3 & 18.2 & 448 & 238.7 & 63.0 & 442 \\ 
  Olsztyn & 79.5 & 18.7 & 959 & 83.3 & 27.2 & 967 & 76.4 & 25.9 & 891 & 83.6 & 22.9 & 878 & 82.5 & 20.2 & 953 \\ 
  Opole & 55.5 & 16.5 & 786 & 59.8 & 11.8 & 800 & 60.1 & 16.3 & 841 & 56.4 & 15.5 & 782 & 54.1 & 17.6 & 870 \\ 
  Poznań & 268.5 & 41.9 & 893 & 269.0 & 64.9 & 841 & 259.0 & 28.7 & 774 & 278.1 & 65.1 & 719 & 232.6 & 53.5 & 785 \\ 
  Rzeszów & 101.8 & 22.9 & 1050 & 105.3 & 31.8 & 1022 & 107.5 & 30.7 & 1095 & 104.7 & 27.0 & 991 & 91.1 & 33.2 & 1052 \\ 
  Sopot & 6.0 & 17.4 & 16 & 4.2 & 23.8 & 10 & 7.0 & 12.8 & 17 & 3.4 & 10.4 &  8 & 3.5 & 10.7 &  8 \\ 
  Szczecin & 200.5 & 37.1 & 1147 & 196.2 & 20.5 & 1055 & 205.0 & 14.0 & 951 & 189.3 & 25.8 & 759 & 191.1 & 9.2 & 671 \\ 
  Toruń & 91.5 & 12.8 & 1176 & 98.8 & 10.0 & 1211 & 92.3 & 14.7 & 1055 & 95.2 & 14.4 & 1039 & 95.8 & 11.4 & 916 \\ 
  Warsaw & 968.2 & 26.8 & 1827 & 978.5 & 63.5 & 1677 & 899.7 & 28.9 & 1660 & 944.8 & 19.4 & 1445 & 909.2 & 50.7 & 1454 \\ 
  Wrocław & 320.9 & 61.2 & 858 & 350.5 & 49.1 & 1006 & 343.2 & 67.8 & 977 & 342.2 & 92.4 & 833 & 314.6 & 97.0 & 805 \\ 
  Zielona Góra & 63.9 & 11.0 & 856 & 65.9 & 10.9 & 810 & 65.2 & 16.0 & 784 & 65.9 & 12.3 & 790 & 68.5 & 13.9 & 811 \\ 
   \hline
\end{tabular}
\end{table}
\end{landscape}

\end{document}